\documentclass[twocolumn,prb]{revtex4}

\usepackage{graphicx}

\def\be{\begin{eqnarray}}
\def\ee{\end{eqnarray}}

\begin{document}

\title{Topological Quantum Compiling}

\author{L. Hormozi, G. Zikos, N.~E. Bonesteel}
\affiliation{Department of Physics and National High Magnetic Field
Laboratory, Florida State University, Tallahassee, Florida 32310}
\author{S.~H. Simon}
\affiliation{Bell Laboratories, Lucent Technologies, Murray Hill,
New Jersey 07974}

\begin{abstract}
A method for compiling quantum algorithms into specific braiding
patterns for nonabelian quasiparticles described by the so-called
Fibonacci anyon model is developed. The method is based on the
observation that a universal set of quantum gates acting on qubits
encoded using triplets of these quasiparticles can be built entirely
out of three-stranded braids (three-braids).  These three-braids can
then be efficiently compiled and improved to any required accuracy
using the Solovay-Kitaev algorithm.
\end{abstract}



\maketitle

\section{Introduction}

The requirements for realizing a fully functioning quantum computer
are daunting. There must be a scalable system of qubits which can be
initialized and individually measured.  It must be possible to enact a
universal set of quantum gates on these qubits. And all this must be
done with sufficient accuracy so that quantum error correction can be
used to prevent decoherence from spoiling any computation.

The problems of error and decoherence are particularly difficult
ones for any proposed quantum computer.  While the states of
classical computers are typically stored in macroscopic degrees of
freedom which have a built-in redundancy and thus are resistant to
errors, building similar redundancy into quantum states is less
natural.  To protect quantum information it is necessary to encode
it using quantum error correcting code states.\cite{shor,steane}
These states are highly entangled, and have the property that code
states corresponding to different logical qubit states can be
distinguished from one another only by global (``topological")
measurements. Unlike states whose macroscopic degrees of freedom
are effectively classical (think of the magnetic moment of a small
part of a hard drive), such highly entangled ``topologically
degenerate'' states do not typically emerge as the ground states
of physical Hamiltonians. One route to fault-tolerant quantum
computation is therefore to build the encoding and fault-tolerant
gate protocols into the ``software" of the quantum
computer.\cite{aharonov}

A remarkable recent development in the theory of quantum computation
which directly addresses these issues has been the realization that
certain exotic states of matter in two space dimensions, so-called
nonabelian states, may provide a natural medium for storing and
manipulating quantum
information.\cite{kitaev,freedman1,freedman2,freedmanbams}  In these
states, localized quasiparticle excitations have quantum numbers
which are in some ways similar to ordinary spin quantum numbers.
However, unlike ordinary spins, the quantum information associated
with these quantum numbers is stored globally, throughout the entire
system, and so is intrinsically protected against decoherence.
Furthermore, these quasiparticles satisfy so-called nonabelian
statistics. This means that when two quasiparticles are
adiabatically moved around one another, while being kept
sufficiently far apart, the action on the Hilbert space is
represented by a unitary matrix which depends only on the topology
of the path used to carry out the exchange. Topological quantum
computation can then be carried out by moving quasiparticles around
one another in two space dimensions.\cite{kitaev,freedman1} The
quasiparticle world-lines form topologically nontrivial braids in
three (= 2 + 1) dimensional space-time, and because these braids are
topologically robust (i.e., they cannot be unbraided without cutting
one of the strands) the resulting computation is protected against
error.

Nonabelian states are expected to arise in a variety of quantum
many-body systems, including spin systems,\cite{fendley,nayak,levin}
rotating Bose gases,\cite{cooper} and Josephson junction
arrays.\cite{doucot} Of those states which have actually been
experimentally observed, the most likely to possess nonabelian
quasiparticle excitations are certain fractional quantum Hall
states. Moore and Read\cite{mooreread} were the first to propose
that quasiparticle excitations which obey nonabelian statistics
might exist in the fractional quantum Hall effect. Their proposal
was based on the observation that the conformal blocks associated
with correlation functions in the conformal field theory describing
the two-dimensional Ising model could be interpreted as quantum Hall
wave functions. These wave functions describe both the ground state
of a half-filled Landau level of spin-polarized electrons, as well
as states with some number of fractionally charged quasihole
excitations (charge = $e/4$).  The particular ground state this
construction produces, the so-called Pfaffian, or Moore-Read state,
is considered the most likely candidate for the observed fractional
quantum Hall state at Landau level filling fraction $\nu = 5/2$
($\nu = 1/2$ in the second Landau level).\cite{morf,rezayi00}

In this conformal field theory construction, states with four or
more quasiholes present correspond to finite-dimensional conformal
blocks, and so the corresponding wave functions form a
finite-dimensional Hilbert space. The monodromy --- or braiding
properties --- of these conformal blocks are then assumed to
describe the unitary transformations acting on the Hilbert space
produced by adiabatically braiding quasiholes around one
another.\cite{mooreread}  Explicit wave functions for these states
were worked out in Ref.~\onlinecite{nayakwilczek}, and the
nonabelian braiding properties have been verified numerically in
Ref.~\onlinecite{simon03}. In an alternate approach, the
Moore-Read state can be viewed as a composite fermion
superconductor in a so-called ``weak pairing" $p_x + i p_y$
phase.\cite{readgreen} In this description, the finite-dimensional
Hilbert space arises from zero energy solutions of the
Bogoliubov-DeGennes equations in the presence of
vortices,\cite{readgreen} and the vortices themselves are
nonabelian quasiholes whose braiding properties have been shown to
agree with the conformal field theory result.\cite{ivanov,stern}
Recently, a number of experiments have been proposed to directly
probe the nonabelian nature of these
excitations.\cite{dassarma05,stern06,bonderson06,hou06}

Unfortunately, the braiding properties of quasihole excitations in
the Moore-Read state are not sufficiently rich to carry out purely
topological quantum computation, although ``partially'' topological
quantum computation using a mixture of topological and
non-topological gates has been shown to be
possible.\cite{bravyi,freedman3} However, Read and
Rezayi\cite{readrezayi} have shown that the Moore-Read state is just
one of a sequence of states labeled by an index $k$ corresponding to
electrons at filling fractions $\nu = k/(2+k)$, with $k=1$
corresponding to the $\nu=1/3$ Laughlin state and $k=2$ to the
Moore-Read state.  The wavefunctions for these states can be written
as correlation functions in the $Z_k$ parafermion conformal field
theory,\cite{readrezayi} and the braiding properties of the
quasihole excitations were worked out in detail in
Ref.~\onlinecite{slingerland01}. There it was shown that the
quasiholes are described by the $SU(2)_k$ Chern-Simons-Witten (CSW)
theories, up to overall abelian phase factors which are irrelevant
for quantum computation.  More recently, explicit quasihole wave
functions have been worked out for the $k = 3$ Read-Reazyi
state,\cite{ardonne} with results consistent with the predicted
$SU(2)_3$ braiding properties. The elementary braiding matrices for
the $SU(2)_k$ CSW theory for $k = 3$ and $k \ge 5$ have been shown
to be sufficiently rich to carry out universal quantum computation,
in the sense that any desired unitary operation on the Hilbert space
of $N$ quasiparticles, with $N \ge 3$ for $k \ge 3$,  $k \neq 4, 8$,
and $N \ge 4$ for $k = 8$,  can be approximated to any desired
accuracy by a braid.\cite{freedman1,freedman2}

The main purpose of this paper is to give an efficient method for
determining braids which can be used to carry out a universal set
of a quantum gates (i.e. single-qubit rotations and controlled-NOT
gates) on encoded qubits for the case $k=3$, thought to be
physically relevant for the experimentally observed\cite{xia} $\nu
= 12/5$ fractional quantum Hall effect\cite{readrezayi,rezayiread}
($\nu = 12/5$ corresponds to $\nu = 2/5$ in the second Landau
level, and this is the particle-hole conjugate of $\nu = 3/5$
corresponding to $k=3$). We refer to the process of finding such
braids as ``topological quantum compiling'' since these braids can
then be used to translate a given quantum algorithm into the
``machine code'' of a topological quantum computer. This is
analogous to the action of an ordinary compiler which translates
instructions written in a high level programming language into the
machine code of a classical computer.

It should be noted that the proof of universality for $SU(2)_3$
quasiparticles is a constructive one,\cite{freedman1,freedman2} and
therefore, as a matter of principle, it provides a prescription for
compiling quantum gates into braids. However, in practice, for
two-qubit gates (such as controlled-NOT) this prescription, if
followed straightforwardly, is prohibitively difficult to carry out,
primarily because it involves searching the space of braids with six
or more strands. We address this difficulty by dividing our
two-qubit gate constructions into a series of smaller constructions,
each of which only involves searching the space of three-stranded
braids (three-braids). The required three-braids then {\it can} be
found efficiently and used to construct the desired two-qubit gates.
This ``divide and conquer" approach does not, in general, yield the
most accurate braid of a given length which approximates a desired
quantum gate. However, we believe that it does yield the most
accurate (or at least among the most accurate) braids which can be
obtained for a given fixed amount of classical computing power.

This paper is organized as follows.  In Sec.~II we review the
basic properties of the $SU(2)_k$ Hilbert space, and show that the
case $SU(2)_3$ is, for our purposes, equivalent to the case
$SO(3)_3$ -- the so-called Fibonacci anyon model.  Section III
then presents a quick review of the mathematical machinery needed
to compute with Fibonacci anyons.  In Sec.~IV we outline how, in
principle, these particles can be used to encode qubits suitable
for quantum computation. Section V then describes how to find
braiding patterns for three Fibonacci anyons which can be used to
carry out any allowed operation on the Hilbert space of these
quasiparticles to any desired accuracy, thus effectively
implementing the procedure given in Ref.~\onlinecite{freedman1}
for carrying out single-qubit rotations. In Sec.~VI we discuss the
more difficult case of two-qubit gates, and give two classes of
explicit gate constructions --- one, first discussed by the
authors in Ref.~\onlinecite{bonesteel05}, in which a pair of
quasiparticles from one qubit is ``woven'' through the
quasiparticles in the second qubit, and another, presented here
for the first time, in which only a single quasiparticle is woven.
Finally, in Sec.~VII we address the question of to what extent the
constructions we find are special to the $k=3$ case, and in
Sec.~VIII we summarize our results.

\section{Fusion Rules and Hilbert Space}

Consider a system with quasiparticle excitations described by the
$SU(2)_k$ CSW theory.  It is convenient to describe the properties
of this system using the so-called quantum group
language.\cite{slingerland01}  The relevant quantum groups are
``deformed'' versions of the representation theory of $SU(2)$,
i.e. the theory of ordinary spin, and much of the intuition for
thinking about ordinary spin can be carried over to the quantum
group case.

\begin{figure}[t]
\includegraphics[scale=.5]{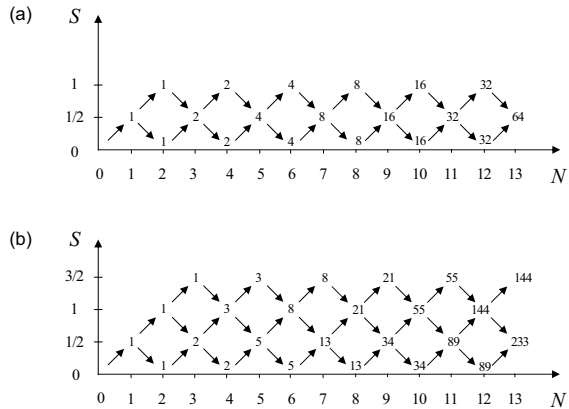}
\caption{Bratteli diagrams for $SU(2)_k$ for (a) $k=2$ and (b)
$k=3$. Here $N$ is the number of $q$-spin 1/2 quasiparticles and
$S$ is the total $q$-spin of those quasiparticles.  The number at
a given $(N,S)$ vertex of each diagram indicates the number of
paths to that vertex starting from the $(0,0)$ point.   This
number gives the dimensionality of the Hilbert space of $N$
$q$-spin 1/2 quasiparticles with total $q$-spin $S$.}
\label{bratteli}
\end{figure}

In the quantum group description of an $SU(2)_k$ CSW theory, each
quasiparticle has a half-integer $q$-deformed spin ($q$-spin)
quantum number. Just as for ordinary spin, there are rules for
combining $q$-spin known as fusion rules. The fusion rules for the
$SU(2)_k$ theory are similar to the usual triangle rule for adding
ordinary spin, except that they are truncated so that there are no
states with total $q$-spin $> k/2$. Specifically, the fusion rules
for the level $k$ theory are,\cite{fuchsbook}
\begin{eqnarray}
s_1 \otimes s_2 &=& |s_1 - s_2| \oplus |s_1 - s_2| + 1
\oplus \ldots \nonumber \\
&&\ \ \ \ \ \ldots  \oplus \min(s_1 + s_2, k - s_1 -
s_2).\label{fusion}
\end{eqnarray}
Note that in the quantum group description of nonabelian anyons,
states are distinguished only by their total $q$-spin quantum
numbers. The $q$-deformed analogs of the $S_z$ quantum numbers are
physically irrelevant --- there is no degeneracy associated with
them, and they play no role in any computation involving
braiding.\cite{slingerland01} The situation is somewhat analogous
to that of a collection of ordinary spin-1/2 particles in which
the only allowed operations, including measurement, are
rotationally invariant and hence independent of $S_z$, as is the
case in exchange-based quantum computation.\cite{divincenzo}

The fusion rules of the $SU(2)_k$ theory fix the structure of the
Hilbert space of the system. For a collection of quasiparticles with
$q$-spin 1/2, a useful way to visualize this Hilbert space is in
terms of its so-called Bratteli diagram. This diagram shows the
different fusion paths for $N$ $q$-spin 1/2 quasiparticles in which
these quasiparticles are fused, one at a time, going from left to
right in the diagram. Bratteli diagrams for the cases $k=2$ and
$k=3$ are shown in Fig.~\ref{bratteli}.

The dimensionality of the Hilbert space for $N$ $q$-spin 1/2
quasiparticles with total $q$-spin $S$ can be determined by
counting the number of paths in the Bratteli diagram from the
origin to the point $(N,S)$.  The results of this path counting
are also shown in Fig.~\ref{bratteli}, where one can see the
well-known $2^{N/2 -1}$ Hilbert space degeneracy for the $k=2$
(Moore-Read) case,\cite{mooreread,nayakwilczek} and the Fibonnaci
degeneracy for the $k=3$ case.\cite{readrezayi}

In this paper we will focus on the $k=3$ case, which is the lowest
$k$ value for which $SU(2)_k$ nonabelian anyons are universal for
quantum computation.\cite{freedman1,freedman2}  In fact, we will
show that two-qubit gates are particularly simple for this case.
Before proceeding, it is convenient to introduce an important
property of the $SU(2)_3$ theory, namely that the braiding
properties of $q$-spin 1/2 quasiparticles are the same as those with
$q$-spin 1 (up to an overall abelian phase which is irrelevant for
topological quantum computation).  This is a useful observation
because the theory of $q$-spin 1 quasiparticles in $SU(2)_3$ is
equivalent to $SO(3)_3$, a theory also known as the Fibonacci anyon
theory\cite{preskillnotes,kuperberg} --- a particularly simple
theory with only two possible values of $q$-spin, 0 and 1, for which
the fusion rules are
\begin{eqnarray}
0 \otimes 0 = 0,\ \ \ \ \ 0\otimes 1 = 1 \otimes 0 = 1,\ \ \ \ \ 1
\otimes 1 = 0 \oplus 1. \label{fibfusion}
\end{eqnarray}

\begin{figure}[t]
\centerline{\includegraphics[scale=.22,angle = 0]{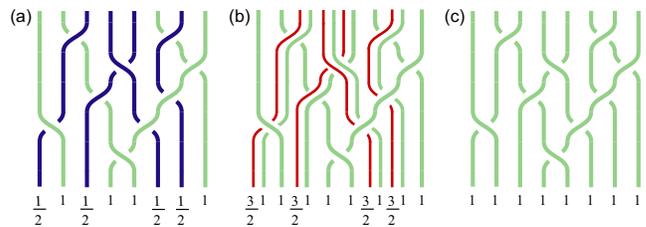}}
\caption{(color online). Graphical proof of the equivalence of
braiding $q$-spin-1/2 and $q$-spin-1 objects for $SU(2)_3$.  Part
(a) shows a braiding pattern for a collection of objects, some
having $q$-spin 1/2 and some having $q$-spin 1.  Part (b) shows the
same braiding pattern but with the $q$-spin-1/2 objects represented
by $q$-spin 1 objects fused with $q$-spin-3/2 objects, which, for
$SU(2)_3$, has a unique fusion channel.  Finally, part (c) shows the
same braid with the $q$-spin-3/2 objects removed.  Because these
$q$-spin-3/2 objects are effectively abelian for $SU(2)_3$, removing
them from the braid will only result in an overall phase factor
which will be irrelevant for quantum computing.} \label{blob}
\end{figure}

Here we give a rough proof of this equivalence. This proof is
based on the fact that for $k=3$ the fusion rules involving
$q$-spin 3/2 quasiparticles take the following simple form
\begin{eqnarray}
\frac{3}{2} \otimes s &=& \frac{3}{2}-s. \label{3/2fusion}
\end{eqnarray}
The key observation is that since for $k=3$ the highest possible
$q$-spin is 3/2, when fusing a $q$-spin-3/2 object with any other
object (here we use the term object to describe either a single
quasiparticle or a group of quasiparticles viewed as a single
composite entity), the Hilbert space dimensionality does not grow.
This implies that moving a $q$-spin-3/2 object around other objects
can, at most, produce an overall abelian phase factor. While this
phase factor may be important physically, particularly in
determining the outcome of interference experiments involving
nonabelian
quasiparticles,\cite{dassarma05,stern06,bonderson06,hou06} it is
irrelevant for quantum computing, and thus does not matter when
determining braids which correspond to a given computation. Because
(\ref{3/2fusion}) implies that a $q$-spin-1/2 object can be viewed
as the result of fusing a $q$-spin-1 object with a $q$-spin-3/2
object, it follows that the braid matrices for $q$-spin-1/2 objects
are the same as that for $q$-spin-1 objects up to an overall phase
(as can be explicitly checked).

In fact, based on this argument we can make a stronger statement.
Imagine a collection of $SU(2)_3$ objects which each have either
$q$-spin 1 or $q$-spin 1/2. It is then possible to carry out
topological quantum computation, {\it even if we do not know which
objects have $q$-spin 1 and which have $q$-spin 1/2}. The proof is
illustrated in Fig.~\ref{blob}. Figure \ref{blob}(a) shows a
braiding pattern for a collection of objects, some of which have
$q$-spin 1/2 and some of which have $q$-spin 1. Fig.~\ref{blob}(b)
then shows the same braiding pattern, but now all objects with
$q$-spin 1/2 are represented by objects with $q$-spin 1 fused to
objects with $q$-spin 3/2. Because, as noted above, the $q$-spin 3/2
objects have trivial (abelian) braiding properties, the unitary
transformation produced by this braid is the same, up to an overall
abelian phase, as that produced by braiding nothing but $q$-spin 1
objects, as shown in Fig.~\ref{blob}(c). It follows that provided
one can measure whether the total $q$-spin of some object belongs to
the class $1 \equiv \{1, 1/2\}$ or the class $0 \equiv \{0, 3/2\}$
--- something which should, in principle, be possible by performing
interference experiments as described in Refs.~\onlinecite{sb} and
\onlinecite{slingerland} --- then quantum computation is possible,
even if we do not know which objects have $q$-spin 1/2 and which
have $q$-spin 1.

\section{Fibonacci Anyon Basics}

Having reduced the problem of compiling braids for $SU(2)_3$ to
compiling braids for $SO(3)_3$, i.e. Fibonacci anyons, it is useful
for what follows to give more details about the mathematical
structure associated with these quasiparticles.  For an excellent
review of this topic see Ref.~\onlinecite{preskillnotes}, and for
the mathematics of nonabelian particles in general see
Ref.~\onlinecite{kitaev2}.

Note that for the rest of this paper, except for Sec. VII, it should
be understood that each quasiparticle is a $q$-spin 1 Fibonacci
anyon.  It should also be understood that from the point of view of
their nonabelian properties quasihole excitations are also $q$-spin
1 Fibonacci anyons, even though they have opposite electric charge
and give opposite abelian phase factors when braided.  Because it is
the nonabelian properties which are relevant for topological quantum
computation, for our purposes quasiparticles and quasiholes can be
viewed as identical nonabelian particles. Unless it is important to
distinguish between the two (as when we discuss creating and fusing
quasiparticles and quasiholes in Sec.~IV) we will simply use the
terms quasiparticle or Fibonacci anyon to refer to either
excitation.

\begin{figure}[t]
\centerline{\includegraphics[scale=.45,angle =
0]{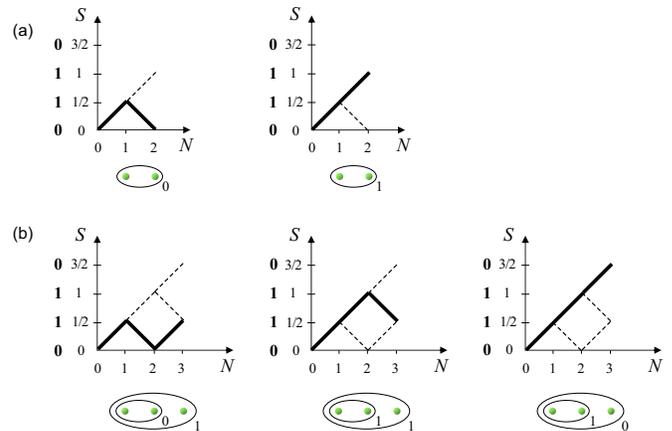}} \caption{(color online). Basis states for the
Hilbert space of (a) two and (b) three Fibonacci anyons.  $SU(2)_3$
Bratteli diagrams showing fusion paths corresponding to the basis
states for the Hilbert space of two and three $q$-spin 1/2
quasiparticles are shown. The $q$-spin axes on these diagrams are
labeled both by the $SU(2)_3$ $q$-spin quantum numbers 0, 1/2, 1 and
3/2 and, to the left of these in bold, the corresponding Fibonacci
$q$-spin quantum numbers $0 \equiv \{0,3/2\}$ and $1 \equiv
\{1/2,1\}$. Beneath each Bratteli diagram the same state is
represented using a notation in which dots correspond to Fibonacci
anyons, and groups of Fibonacci anyons enclosed in ovals labeled by
$q$-spin quantum numbers are in the corresponding $q$-spin
eigenstates.} \label{notation}
\end{figure}

Figure \ref{notation} establishes some of the notation for
representing Fibonacci anyons which will be used in the rest of
the paper. This figure shows $SU(2)_3$ Bratteli diagrams in which
the $q$-spin axis is labeled both by the $SU(2)_3$ $q$-spin
quantum numbers and, in boldface, the corresponding Fibonacci
$q$-spin quantum numbers, i.e. 0 for $\{0,3/2\}$ and 1 for
$\{1/2,1\}$. In Fig.~\ref{notation}(a) Bratteli diagrams showing
fusion paths corresponding to two basis states spanning the
two-dimensional Hilbert space of two Fibonacci anyons are shown.
Beneath each Bratteli diagram an alternate representation of the
corresponding state is also shown. In this representation dots
correspond to Fibonacci anyons and ovals enclose collections of
Fibonacci anyons which are in $q$-spin eigenstates whenever the
oval is labeled by a total $q$-spin quantum number. (Note: If the
oval is not labeled, it should be understood that the enclosed
quasiparticles may not be in a $q$-spin eigenstate).

In the text we will use the notation $\bullet$ to represent a
Fibonacci anyon, and the ovals will be represented by parentheses.
In this notation, the two states shown in Fig.~\ref{notation}(a)
are denoted $(\bullet,\bullet)_0$ and $(\bullet,\bullet)_1$.

Fig.~\ref{notation}(b) shows Bratteli diagram, again with both
$SU(2)_3$ and Fibonacci quantum numbers, with fusion paths which
this time correspond to three basis states of the three-dimensional
Hilbert space of three Fibonacci anyons.  Beneath these diagrams the
``oval" representations of these three states are also shown, which
in the text will be represented $((\bullet,\bullet)_0,\bullet)_1$,
$((\bullet,\bullet)_1,\bullet)_1$ and
$((\bullet,\bullet)_1,\bullet)_0$.

In addition to fusion rules, all theories of nonabelian anyons
possess additional mathematical structure which allows one to
calculate the result of any braiding operation.  This structure is
characterized by the $F$ (fusion) and $R$ (rotation)
matrices.\cite{preskillnotes,kitaev2,mooreseiberg}

To define the $F$ matrix, note that the Hilbert space of three
Fibonacci anyons is spanned both by the three states labeled
$((\bullet,\bullet)_a,\bullet)_c$, and the three states labeled
$(\bullet,(\bullet,\bullet)_b)_c$.  The $F$ matrix is the unitary
transformation which maps one of these bases to the other,
\begin{eqnarray}
\left(\bullet,\left(\bullet,\bullet\right)_a \right)_c = \sum_{b}
F^{c}_{ab} \left(\left(\bullet,\bullet\right)_b,\bullet\right)_c,
\end{eqnarray}
and has the form
\be F
= \left(\begin{array}{cc|c}\tau&\sqrt{\tau}& \\\sqrt{\tau}&-\tau& \\
\hline  & & 1\end{array}\right), \label{fmatrix}\ee
where $\tau = (\sqrt{5}-1)/2$ is the inverse of the golden mean. In
this matrix the upper left 2$\times$2 block, $F^1_{ab}$, acts on the
two-dimensional total $q$-spin 1 sector of the three-quasiparticle
Hilbert space and the lower right matrix element, $F^0_{11} = 1$,
acts on the unique total $q$-spin 0 state. Note that this $F$ matrix
can be applied to any three objects which each have $q$-spin 1,
where each object can consist of more than one Fibonacci anyon.
Furthermore, if one considers three objects for which one or more of
the objects has $q$-spin 0, then the state of these objects is
uniquely determined by the total $q$-spin of all three, and in this
case the $F$ matrix is trivially the identity. Thus, for the case of
Fibonacci anyons, the matrix (\ref{fmatrix}) is all that is needed
to make arbitrary basis changes for any number of Fibonacci anyons.

The $R$ matrix gives the phase factor produced when two Fibonacci
anyons are moved around one another with a certain sense.  One can
think of these phase factors as the $q$-deformed versions of the
$-1$ or $+1$ phase factors one obtains when interchanging two
ordinary spin-1/2 quasiparticles when they are in a singlet or
triplet state, respectively. This phase factor depends on the
overall $q$-spin of the two quasiparticles involved in the exchange,
so for Fibonacci anyons there are two such phase factors which are
summarized in the $R$ matrix,
\begin{eqnarray}
R = \left( \begin{array}{cc} e^{-i4 \pi/5} & 0 \\ 0 & e^{i 3\pi/5}
\end{array}\right).
\end{eqnarray}
Here the upper left and lower right matrix elements are,
respectively, the phase factor that two Fibonacci anyons acquire
if they are interchanged in a clockwise sense when they have total
$q$-spin 0 or $q$-spin 1.  Again, this matrix also applies if we
exchange two objects that both have total $q$-spin 1, even if
these objects consist of more than one Fibonacci anyon.  And if
one or both objects has $q$-spin 0, the result of this interchange
is the identity.  Again we emphasize that in the $k=3$ Read-Rezayi
state, there will be additional abelian phases present, which may
have physical consequences for some experiments, but which will be
irrelevant for topological quantum computation.

Typically the sequence of $F$ and $R$ matrices used to compute the
unitary operation produced by a given braid is not unique.  To
guarantee that the result of any such computation is independent of
this sequence, the $F$ and $R$ matrices must satisfy certain
consistency conditions. These consistency conditions, the so-called
pentagon and hexagon
equations,\cite{preskillnotes,kitaev2,mooreseiberg} are highly
restrictive, and, in fact, for the case of Fibonacci anyons
essentially fix the $F$ and $R$ matrices to have the forms given
above (up to a choice of chirality, and Abelian phase factors which
are again irrelevant to our purposes here).\cite{preskillnotes}

Finally, we point out an obvious, but important, consequence of
the structure of the $F$ and $R$ matrices. When interchanging any
two quasiparticles which are part of a larger set of
quasiparticles with a well-defined total $q$-spin quantum number,
this total $q$-spin quantum number will not change.

\begin{figure}[t]
\includegraphics[scale=.5]{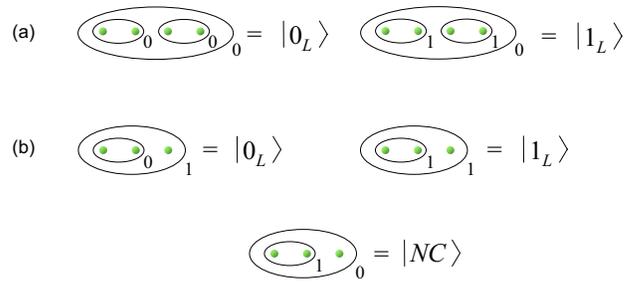}
\caption{(color online). (a) Four-quasparticle and (b)
three-quasiparticle qubit encodings for Fibonacci anyons. Part (a)
shows two states which span the Hilbert space of four quasiparticles
with total $q$-spin 0 which can be used as the logical $|0_L\rangle$
and $|1_L\rangle$ states of a qubit.  Part (b) shows two states
spanning the Hilbert space of three quasiparticles with total
$q$-spin 1 which can also be used as logical qubit states
$|0_L\rangle$ and $|1_L\rangle$. This three-quasiparticle qubit can
be obtained by removing the rightmost quasiparticle from the two
states shown in (a). The third state shown in Part (b), labeled
$|NC\rangle$ for ``noncomputational", is the unique state of three
quasiparticles which has total $q$-spin 0.} \label{qubits}
\end{figure}

\section{Qubit Encoding and General Computation Scheme}

Before proceeding, it will be useful to have a specific scheme in
mind for how one might actually carry out topological quantum
computation with Fibonacci anyons.  Here we follow the scheme
outlined in Ref.~\onlinecite{freedmanbams}, which, for
completeness, we briefly review below.

The computer can be initialized by pulling quasiparticle-quasihole
pairs out of the ``vacuum", (by vacuum we mean the ground state of
the $k=3$ Read-Rezayi state or any other state which supports
Fibonacci anyon excitations). Each such pair will consist of two
$q$-spin 1 excitations in a state with total $q$-spin 0, i.e. the
state $(\bullet,\bullet)_0$. In principle, this pair can also
exist in a state with total $q$-spin 1, provided there are other
quasiparticles present to ensure the total $q$-spin of the system
is 0, so one can imagine using this pair as a qubit. However, it
is impossible to carry out arbitrary single-qubit operations by
braiding only the two quasiparticles forming such a qubit --- this
braiding never changes the total $q$-spin of the pair, and so only
generates rotations about the $z$-axis in the qubit space.

For this reason it is convenient to encode qubits using more than
two Fibonacci anyons.  Thus, to create a qubit, {\it two}
quasiparticle-quasihole pairs can be pulled out of the vacuum. The
resulting state is then
$((\bullet,\bullet)_0,(\bullet,\bullet)_0)_0$ which again has
total $q$-spin 0. The Hilbert space of four Fibonacci anyons with
total $q$-spin 0 is two dimensional, with basis states, which we
can take as logical qubit states,  $|0_L\rangle =
((\bullet,\bullet)_0,(\bullet,\bullet)_0)_0$ and $|1_L\rangle =
((\bullet,\bullet)_1,(\bullet,\bullet)_1)_0$, (see
Fig~\ref{qubits}(a)). The state of such a four-quasiparticle qubit
is determined by the total $q$-spin of either the rightmost or
leftmost pair of quasiparticles. Note that the fusion rules
(\ref{fibfusion}) imply that the total $q$-spin of these two pairs
must be the same because the total $q$-spin of all four
quasiparticles is 0.

For this encoding, in addition to the two-dimensional
computational qubit space of four quasiparticles with total
$q$-spin 0, there is a three-dimensional {\it noncomputational}
Hilbert space of states with total $q$-spin 1 spanned by the
states $((\bullet,\bullet)_0,(\bullet,\bullet)_1)_1$,
$((\bullet,\bullet)_1,(\bullet,\bullet)_0)_1$ and
$((\bullet,\bullet)_1,(\bullet,\bullet)_1)_1$.  When carrying out
topological quantum computation it is crucial to avoid transitions
into this noncomputational space.

Fortunately, single-qubit rotations can be carried out by braiding
quasiparticles within a given qubit and, as discussed in Sec.~III,
such operations will not change the total $q$-spin of the four
quasiparticles involved.  Single-qubit operations can therefore be
carried out without any undesirable transitions out of the encoded
computational qubit space.

Two-qubit gates, however, will require braiding quasiparticles from
different qubits around one another.  This will in general lead to
transitions out of the encoded qubit space. Nevertheless, given the
so-called "density" result of Ref.~\onlinecite{freedman2} it is
known that, as a matter of principle, one can always find two-qubit
braiding patterns which will entangle the two qubits, and also stay
within the computational space to whatever accuracy is required for
a given computation. The main purpose of this paper is to show how
such braiding patterns can be efficiently found.

Note that the action of braiding the two leftmost quasiparticles in
a four-quasiparticle qubit (referring to Fig.~\ref{qubits}(a)) is
equivalent to that of braiding the two rightmost quasiparticles with
the same sense. This is because as long as we are in the
computational qubit space both the leftmost and rightmost
quasiparticle pairs must have the same total $q$-spin, and so
interchanging either pair will result in the same phase factor from
the $R$ matrix. It is therefore not necessary to braid all four
quasiparticles to carry out single-qubit rotations --- one need only
braid three.

In fact, one may consider qubits encoded using only three
quasiparticles with total $q$-spin 1, as originally proposed in
Ref.~\onlinecite{freedman1}. Such qubits can be initialized by first
creating a four-quasiparticle qubit in the state $|0_L\rangle$, as
outlined above, and then simply removing one of the quasiparticles.
In this three-quasiparticle encoding, shown in Fig.~\ref{qubits}(b),
the logical qubit states can be taken to be $|0_L\rangle =
((\bullet,\bullet)_0,\bullet)_1$ and $|1_L\rangle =
((\bullet,\bullet)_1,\bullet)_1$.  For this encoding there is just a
single noncomputational state $|NC\rangle =
((\bullet,\bullet)_1,\bullet)_0$, also shown in
Fig.~\ref{qubits}(b).  As for the four-quasiparticle qubit, when
carrying out single-qubit rotations by braiding within a
three-quasiparticle qubit the total $q$-spin of the qubit, in this
case 1, remains unchanged and there are no transitions from the
computational qubit space into the state $|NC\rangle$.  However,
just as for four-quasiparticle qubits, when carrying out two-qubit
gates these transitions will in general occur and we must work hard
to avoid them. Henceforth we will refer to these unwanted
transitions as leakage errors.

\begin{figure}[t]
\includegraphics[scale=.13]{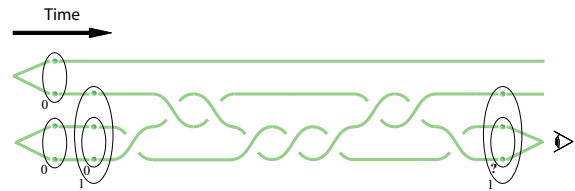}
\caption{(color online). Space-time paths corresponding to the
initialization, manipulation through braiding, and measurement of an
encoded qubit. Two quasiparticle-quasihole pairs are pulled out of
the vacuum, with each pair having total $q$-spin 0. The resulting
state corresponds to a four-quasiparticle qubit in the state
$|0_L\rangle$ (see Fig.~\ref{qubits}(a)). After some braiding, the
qubit is measured by trying to fuse the bottommost pair (in this
case a quasiparticle-quasihole pair). If they fuse back into the
vacuum the result of the measurement is $|0_L\rangle$, otherwise it
is $|1_L\rangle$. Because only the three lower quasiparticles are
braided, the encoded qubit can also be viewed as a
three-quasiparticle qubit (see Fig.~\ref{qubits}(b)) which is
initialized in the state $|0_L\rangle$.} \label{init}
\end{figure}

Note that, because each three-quasiparticle qubit has total $q$-spin
1, when more than one of these qubits is present the state of the
system is not entirely characterized by the ``internal" $q$-spin
quantum numbers which determine the computational qubit states. It
is also necessary to specify the state of what we will refer to as
the ``external fusion space" --- the Hilbert space associated with
fusing the total $q$-spin 1 quantum numbers of each qubit.  When
compiling braids for three-quasiparticle qubits it is crucial that
the operations on the computational qubit space not depend on the
state of this external fusion space --- if they did, these two
spaces would become entangled with one another leading to errors.
Fortunately, we will see that it is indeed possible to find braids
which do not lead to such errors.

For the rest of this paper (except Sec.~VII) we will use this
three-quasiparticle qubit encoding.  It should be noted that any
braid which carries out a desired operation on the computational
space for three-quasiparticle qubits will carry out the same
operation on the computational space of four-quasiparticle qubits,
with one quasiparticle in each qubit acting as a spectator. The
braids we find here can therefore be used for either encoding.

We can now describe how topological quantum computation might
actually proceed, again following Ref.~\onlinecite{freedmanbams}.
A quantum circuit consisting of a sequence of one- and two-qubit
gates which carries out a particular quantum algorithm would first
be translated (or ``compiled'') into a braid by compiling each
individual gate to whatever accuracy is required. Qubits would
then be initialized by pulling quasiparticle-quasihole pairs out
of the ``vacuum". These localized excitations would then be
adiabatically dragged around one another so that their world-lines
trace out a braid in three-dimensional space-time which is
topologically equivalent to the braid compiled from the quantum
algorithm. Finally, individual qubits would be measured by trying
to fuse either the two rightmost or two leftmost excitations
within them (referring to Fig.~\ref{qubits}(a)) for
four-quasiparticle qubits, or just the two leftmost excitations
(referring to Fig.~\ref{qubits}(b)) for three-quasiparticle
qubits. If this pair of excitations consists of a quasiparticle
and a quasihole (and it will always be possible to arrange this),
then, if the total $q$-spin of the pair is 0, it will be possible
for them to fuse back into the ``vacuum". However, if the total
$q$-spin is 1 this will not be possible. The resulting difference
in the charge distribution of the final state would then be
measured to determine if the qubit was in the state $|0_L\rangle$
or $|1_L\rangle$.  Alternatively, as already mentioned in Sec.~II,
interference experiments\cite{sb,slingerland} could be used to
initialize and read out encoded qubits.

As a simple illustration, Fig.~\ref{init} shows a ``computation"
in which a four-quasiparticle qubit (which can also be viewed as a
three-quasiparticle qubit if the top quasiparticle is ignored) is
initialized by pulling quasiparticle-quasihole pairs out of the
vacuum, a single-qubit operation is carried out by braiding within
the qubit, and the final state of the qubit is measured by fusing
a quasiparticle and quasihole together and observing the outcome.

\section{Compiling three-braids and single-qubit gates}

We now focus on the problem of finding braids for three Fibonacci
anyons (three-braids) which approximate any allowed unitary
transformation on the Hilbert space of these quasiparticles.  This
is important not only because it allows one to find braids which
carry out arbitrary single-qubit rotations,\cite{freedman1} but
also because, as will be shown in Sec.~VI, it is possible to
reduce the problem of constructing braids which carry out
two-qubit gates to that of finding a series of three-braids
approximating specific operations.

\begin{figure}
\includegraphics[scale=.15]{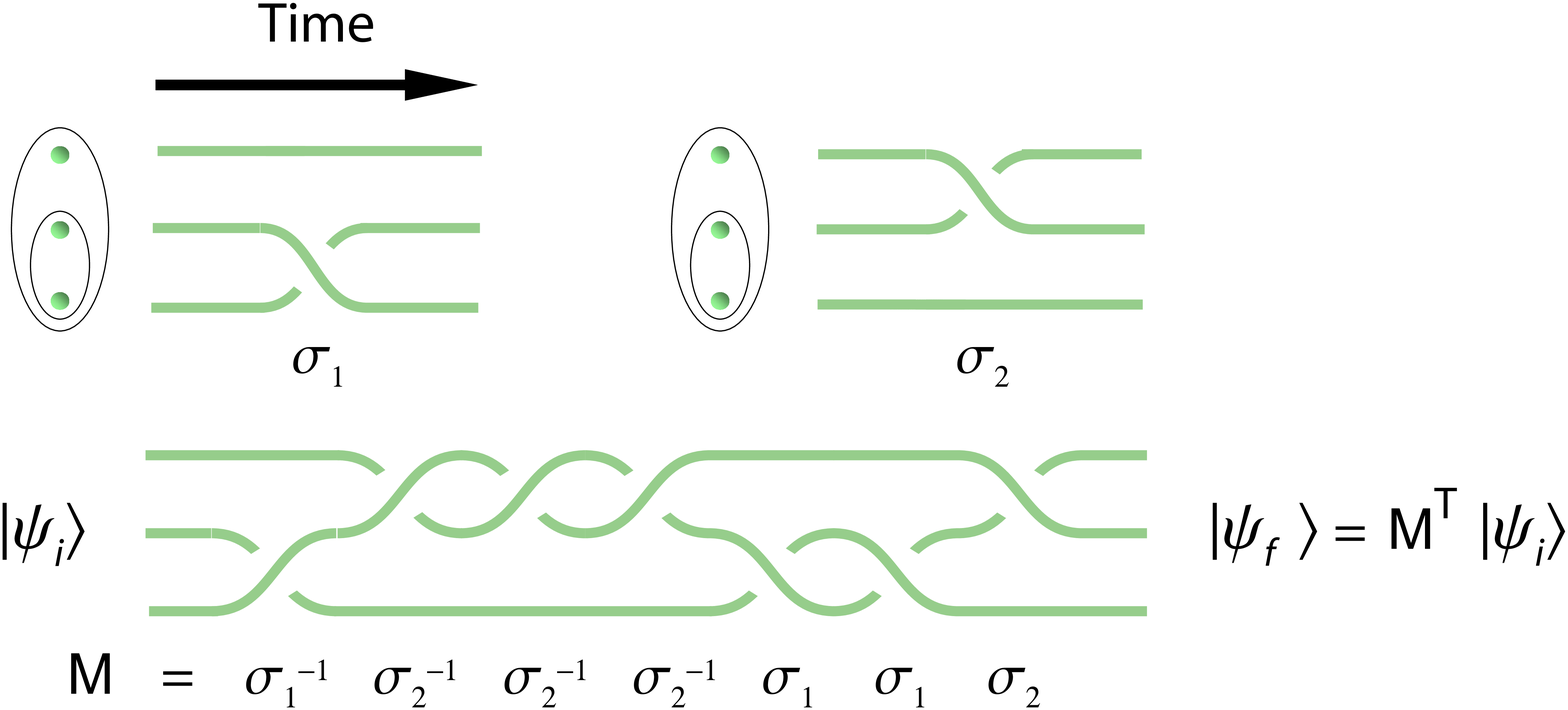}
\caption{(color online). Elementary three-braids and the
decomposition of a general three-braid into a series of elementary
braids.  The unitary operation produced by this braid is computed by
multiplying the corresponding sequence of elementary braid matrices,
$\sigma_1$ and $\sigma_2$ (see text) and their inverses, as shown.
Here the (unlabeled) ovals represent a particular basis choice for
the three-quasiparticle Hilbert space, consistent with that used in
the text. In this and all subsequent figures which show braids,
quasiparticles are aligned vertically, and we adopt the convention
that reading from bottom to top in the figures corresponds to
reading from left to right in expressions such as
$((\bullet,\bullet)_a,\bullet)_c$ in the text.  It should be noted
that these figures are only meant to represent the topology of a
given braid.  In any actual implementation of topological quantum
computation, quasiparticles will certainly not be arranged in a
straight line, and they will have to be kept sufficiently far apart
while being braided to avoid lifting the topological degeneracy. }
\label{basis}
\end{figure}

\subsection{Elementary Braid Matrices}

Using the $F$ and $R$ matrices, it is straightforward to determine
the elementary braiding matrices that act on the three-dimensional
Hilbert space of three Fibonacci anyons.  If, as in
Fig.~\ref{basis}, we take the basis states for the
three-quasiparticle Hilbert space to be the states labeled
$((\bullet,\bullet)_a,\bullet)_c$ then, in the $ac = \{01,11,10\}$
basis, the matrix $\sigma_1$ corresponding to a clockwise
interchange of the two bottommost quasiparticles in the figure (or
leftmost in the $((\bullet,\bullet)_a,\bullet)_c$ representation)
is
\be \sigma_1 = \left(\begin{array}{cc|c}e^{-i4\pi/5}&0&  \\0&
e^{i3\pi/5}&  \\ \hline  & &e^{i3\pi/5}\end{array}\right), \ee
where the upper left 2$\times$2 block acts on the total $q$-spin 1
sector ($|0\rangle_L$ and $|1\rangle_L$) of the three
quasiparticles, and the lower right matrix element is a phase factor
acquired by the $q$-spin 0 state ($|NC\rangle$). This matrix is
easily read off from the $R$ matrix, since the total $q$-spin of the
two quasiparticles being exchanged is well defined in this basis.

To find the matrix $\sigma_2$ corresponding to a clockwise
interchange of the two topmost (or rightmost in the
$((\bullet,\bullet)_a,\bullet)_c$ representation) quasiparticles,
we must first use the $F$ matrix to change bases to one in which
the total $q$-spin of these quasiparticles is well defined. In
this basis, the braiding matrix is simply $\sigma_1$, and so,
after changing back to the original basis, we find
\begin{eqnarray}
\sigma_2 = F^{-1} \sigma_1 F = \left(\begin{array}{cc|c} -\tau
e^{-i\pi/5} & \sqrt{\tau} e^{-i 3\pi/5} &  \\
\sqrt{\tau} e^{-i 3 \pi/5} & -\tau &  \\
\hline & & e^{i 3\pi/5} \end{array}\right) .\label{fsf}
\end{eqnarray}

The unitary transformation corresponding to a given three-braid
can now be computed by representing it as a sequence of elementary
braid operations and multiplying the corresponding sequence of
$\sigma_1$ and $\sigma_2$ matrices and their inverses, as shown in
Fig.~\ref{basis}.

If we are only concerned with single-qubit rotations, then we only
care about the action of these matrices on the encoded qubit space
with total $q$-spin 1, and not the total $q$-spin 0 sector
corresponding to the noncomputational state.  However, in our
two-qubit gate constructions, various three-braids will be
embedded into the braiding patterns of six quasiparticles, and in
this case the action on the full three-dimensional Hilbert space
does matter.

To understand this action note that $\sigma_1$ can be written
\be \sigma_1 = \left(\begin{array}{c|c}
\pm e^{-i\pi/10} \left(\begin{array}{cc} \pm e^{-i7\pi/10} & 0 \\
0 & \pm e^{i7\pi/10}
\end{array}\right) & \\ \hline & e^{i 3\pi/5}
\end{array}\right),~
\ee
where the upper $2\times 2$ block acting on the total $q$-spin 1
sector is an $SU(2)$ matrix, (i.e., a $2\times 2$ unitary matrix
with determinant 1), multiplied by a phase factor of either
$+e^{-i\pi/10}$ or $-e^{-i\pi/10}$, and the lower right matrix
element, $e^{i3\pi/5}$, is the phase acquired by the total $q$-spin
0 state. The phase factor pulled out of the upper $2 \times 2$ block
is only defined up to $\pm 1$ because any $SU(2)$ matrix multiplied
by $-1$ is also an $SU(2)$ matrix.

From (\ref{fsf}) it follows that $\sigma_2$ can be written in a
similar fashion, with the same phase factors. Each clockwise
braiding operation then corresponds to applying an $SU(2)$
operation multiplied by a phase factor of $\pm e^{-i\pi/10}$ to
the $q$-spin 1 sector, while at the same time multiplying the
$q$-spin 0 sector by a phase factor of $e^{i 3\pi/5}$.  Likewise,
each counterclockwise braiding operation corresponds to applying
an $SU(2)$ operation multiplied by a phase factor of $\pm
e^{+i\pi/10}$ to the $q$-spin 1 sector and a phase factor of
$e^{-i3\pi/5}$ to the $q$-spin 0 sector.

We define the winding, $W(B)$, of a given three-braid $B$, to be
the total number of clockwise interchanges minus the total number
of counterclockwise interchanges.  It then follows that the
unitary operation corresponding to an arbitrary braid $B$ can
always be expressed
\be U(B) = \left(\begin{array}{c|c}
\pm e^{-i W(B)\pi/10}\: [SU(2)] & \\
\hline & e^{i3W(B) \pi/5}
\end{array}\right),\label{winding}
\ee
where $[SU(2)]$ indicates an $SU(2)$ matrix. Thus, for a given
three-braid, the phase relation between the total $q$-spin 1 and
total $q$-spin 0 sectors of the corresponding unitary operation is
determined by the winding of the braid.  We will refer to
(\ref{winding}) often in what follows. It tells us precisely what
unitary operations can be approximated by three-braids, and places
useful restrictions on their winding.

\subsection{Weaving and Brute Force Search}

At this point it is convenient to restrict ourselves to a subclass
of braids which we will refer to as weaves.  A weave is any braid
which is topologically equivalent to the space-time paths of some
number of quasiparticles in which only a single quasiparticle
moves. It was shown in Ref.~\onlinecite{simon06} that this
restricted class of braids is universal for quantum computation,
provided the unitary representation of the braid group is dense in
the space of all unitary transformations on the relevant Hilbert
space, which is the case for Fibonacci anyons.

Following Ref.~\onlinecite{simon06} we will borrow some weaving
terminology and refer to the mobile quasiparticle (or collection
of quasiparticles) as the ``weft'' quasiparticle(s) and the static
quasiparticles as the ``warp'' quasiparticles.

One reason for focusing on weaves is that weaving will likely be
easier to accomplish technologically than general braiding. This
is true even if the full computation involves not just weaving a
single quasiparticle, as was proposed in
Ref.~\onlinecite{simon06}, but possibly weaving several
quasiparticles at the same time in different regions of the
computer --- carrying out quantum gates on different qubits in
parallel.

Considering weaves has the added (and more immediate) benefit of
simplifying the problem of numerically searching for three-braids
which approximate desired gates. For the full braid group, even on
just three strands, there is a great deal of redundancy since braids
which are topologically equivalent will yield the same unitary
operation. Weaves, however, naturally provide a unique
representation in which the warp strands are straight, and the weft
weaves around them. There is therefore no trivial ``double
counting'' of topologically equivalent weaves when one does a brute
force numerical search of weaves up to some given length.

The unitary operations performed by weaving three quasiparticles in
which the weft quasiparticle starts and ends in the middle position,
will always have the form
\be U_{\rm weave}(\{n_i\}) = \sigma_1^{n_m} \sigma_2^{n_{m-1}}
\cdots \sigma_1^{n_3} \sigma_2^{n_2} \sigma_1^{n_1}.\label{uweave}
\ee
Here the sequence of exponents $n_2, n_3 \cdots n_{m-1}$ all take
their values from $\{\pm 2, \pm 4\}$, and $n_1$ and $n_m$ can take
the values $\{0, \pm 2, \pm 4 \}$. Because these exponents are all
even, each factor in this sequence takes the weft quasiparticle
all the way around one of the two warp quasiparticles either once
or twice with either a clockwise or counterclockwise sense,
returning it to the middle position.  We allow $n_1$ and $n_m$ to
be 0 to account for the possibility that the initial or final
weaving operations could each be either $\sigma_1^n$ or
$\sigma_2^n$ with $n = \pm 2$ or $\pm 4$.  Note that we need only
consider exponents $n_i$ up to $\pm 4$ (i.e., moving the weft
quasiparticle at most two times around a warp quasiparticle)
because of the fact that $\sigma_i^{10} = 1$ for Fibonacci anyons,
implying, e.g., $\sigma_i^6 = \sigma_i^{-4}$. We define the length
$L$ of such weaves to be equal to the total number of elementary
crossings, thus $L = \sum_{i=1}^m |n_i|$.

We will also consider weaves in which the weft quasiparticle begins
and/or ends at a position other than the middle. These possibilities
can easily be taken into account by multiplying $U_{\rm
weave}(\{n_i\})$, as defined in (\ref{uweave}), by the appropriate
factors of $\sigma_1$ or $\sigma_2$ on the right and/or left.  Thus,
for example, the unitary operation produced by a weave in which the
weft quasiparticle starts in the top position and ends in the middle
position can be written $U_{\rm weave} (\{n_i\}) \sigma_2$, where,
because of the extra factor of $\sigma_2$, the first braiding
operations carried out by this weave will be $\sigma_2^n$ where $n$
is an odd power, $n=\pm 1, \pm 3$ or 5.  This will weave the weft
quasiparticle from the top position to the middle position after
which $U_{\rm weave}$ will simply continue weaving this
quasiparticle eventually ending with it in the middle position.
(Note that by multiplying $U_{\rm weave}$ on the right by
$\sigma_2$, and not $\sigma_2^{-1}$, we are not requiring the
initial elementary braid to be clockwise, since $U_{\rm weave}$ may
have $n_1 = 0$ and $n_2 = -2$ or $-4$ so that the initial $\sigma_2$
is immediately multiplied by $\sigma_2$ to a negative power.)
Similarly, the unitary operation produced by a weave in which the
weft particle starts in the top position and ends in the bottom
position can be written $\sigma_1 U_{\rm weave}(\{n_i\}) \sigma_2$,
and so on.

To find a weave for which the corresponding unitary operation
$U_{\rm weave}(\{n_i\})$ approximates a particular desired unitary
operation, the most straightforward approach is to simply perform
a brute force search over all weaves, i.e. all sequences $\{n_i\}$
as described above, up to a certain length $L$, in order to find
the $U_{\rm weave}(\{n_i\})$ which is closest to the target
operation. Here we will take as a measure of the distance between
two operators $U$ and $V$ the operator norm distance
$\epsilon(U,V) = ||U-V||$ where $||O||$ is the operator norm,
defined to be the square root of the highest eigenvalue of
$O^\dagger O$. Again, if we are interested in fixing the relative
phase of the total $q$-spin 1 and total $q$-spin 0 sectors then we
would restrict the winding of the weaves so that the phases in
(\ref{winding}) match those of the desired target gate.

\begin{figure}
\centerline{\includegraphics[scale=.3,angle=0]{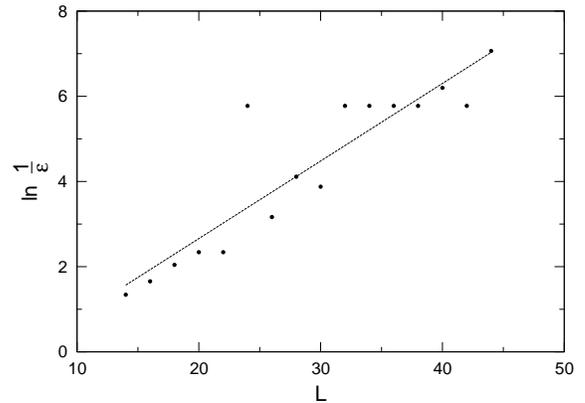}}
\caption{$\ln \frac{1}{\epsilon}$ vs. braid length $L$ for weaves
approximating the gate $iX$.  Here $\epsilon$ is the distance
(defined in terms of operator norm) between $iX$ and the unitary
transformation produced by a weave of length $L$ which best
approximates it. The line is a guide to the eye.} \label{brute}
\end{figure}

For example, imagine our goal is to find a weave which
approximates the unitary operation,
\begin{eqnarray}
i X = \left(\begin{array}{cc|c} 0 & i & \\ i & 0 & \\ \hline  & & 1
\end{array}\right).
\label{ix}
\end{eqnarray}
If the resulting weave were to be used only for a single-qubit
operation, then we would only require that the weave approximate
the upper left $2\times 2$ block of $iX$ up to an overall phase
and we would not care about the phase factor appearing in the
lower right matrix element.  There would then be no constraint on
the winding of the braid.  However, for this example we will
assume that this weave will be used in a two-qubit gate
construction, for which the overall phase and/or the phase
difference between the total $q$-spin 1 and total $q$-spin 0
sectors will matter.

In this case, by comparing $iX$ to (\ref{winding}), we see that
the winding $W$ of any weave approximating $iX$ must satisfy $e^{i
3 \pi W/5} = 1$ or $W = 0$ (modulo 10). Results of a brute force
search over weaves satisfying this winding requirement which
approximate $i X$ are shown in Fig.~\ref{brute}. In this figure,
$\ln \frac{1}{\epsilon}$ is plotted vs. braid length $L$, where
$\epsilon$ is the minimum distance between $U_{\rm weave}$ and
$iX$ for weaves of length $L$.  It is expected that, for any such
brute force search for weaves approximating a generic target
operation, the length should scale with distance according to $L
\sim \log \frac{1}{\epsilon}$, because the number of braids grows
exponentially with $L$.  The results shown in Fig.~\ref{brute} are
consistent with such logarithmic scaling.

All the brute force searches used to find braids in this paper are
straightforward sequential searches, meant mainly to demonstrate
proof of principle.  No doubt more sophisticated brute force search
methods (e.g. bidirectional search) could be used to perform deeper
searches resulting in longer and more accurate braids. Nevertheless,
the exponential growth in the number of braids with $L$ implies that
finding optimal braids by any brute force search method will rapidly
become infeasible as $L$ increases. Fortunately one can still
systematically improve a given braid to any desired accuracy by
applying the Solovay-Kitaev algorithm,\cite{kitaevbook,nielsenbook}
which we now briefly review.

\subsection{Implementation of the Solovay-Kitaev Algorithm for Braids}

The general result of the Solovay-Kitaev theorem tells us that we
can efficiently improve the accuracy of any given braid without the
need to perform exhaustive brute force searches of ever improving
accuracy.\cite{kitaevbook,nielsenbook} The essential ingredient in
this procedure is an $\epsilon$-net --- a discrete set of operators
which in the present case correspond to finite braids up to some
given length, with the property that for any desired unitary
operator there exists an element of the $\epsilon$-net which is
within some given distance $\epsilon_0$ of that operator. Provided
$\epsilon_0$ is sufficiently small, the Solovay-Kitaev algorithm
gives us a clever way to pick a finite number of braid segments out
of the $\epsilon$-net and sew them together so that the resulting
gate will be an approximation to the desired gate with improved
accuracy.

The implementation of the Solovay-Kitaev algorithm we use here
follows closely that described in detail in
Refs.~\onlinecite{harrow} and \onlinecite{dawson}. The first step
of this algorithm is to find a braid which approximates the
desired gate, $U$, by performing a brute force search over the
$\epsilon$-net. Let $U_0$ denote the result of this search. Since
we know that $\epsilon(U_0,U) \le \epsilon_0$ it follows that $C =
U U_0^{-1}$ is an operator which is within a distance $\epsilon_0$
of the identity.

The next step is to decompose $C$ as a group commutator.  This
means that we find two unitary operators $A$ and $B$ for which $C
= A B A^{-1} B^{-1}$.  The unitary operators $A$ and $B$ are
chosen so that their action on the computational qubit space
corresponds to small rotations through the same angle but about
perpendicular axes. For this choice, if $A$ and $B$ are then
approximated by operators $A_0$ and $B_0$ in the $\epsilon$-net,
it can readily be shown that the operator $C_0 = A_0 B_0 A_0^{-1}
B_0^{-1}$, will approximate $C$ to a distance of order
$\epsilon_0^{3/2}$.  It follows that the operator $U_1 = A_0 B_0
A_0^{-1} B_0^{-1} U_0$ is an approximation to $U$ within a
distance $\epsilon_1 \simeq c \epsilon_0^{3/2}$, where $c$ is a
constant which determines the size of the $\epsilon$-net needed to
guarantee an improvement in accuracy.

\begin{figure}
\includegraphics[scale=.8]{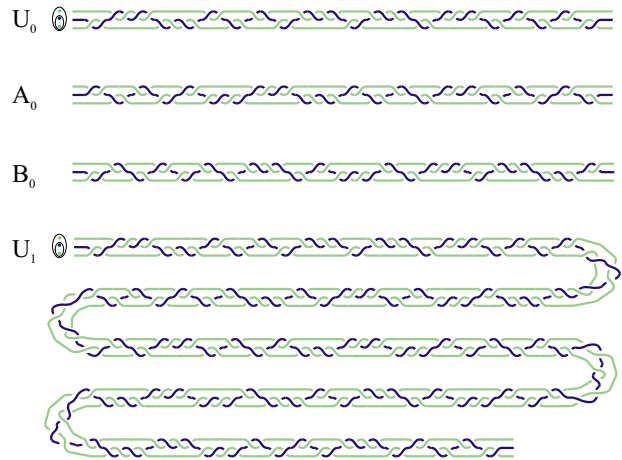}
\caption{(color online). One iteration of the Solovay-Kitaev
algorithm applied to finding a braid which approximates the
operation $U = iX$.  The braid $U_0$ is the result of a brute force
search over weaves up to length 44 which best approximates the
desired gate $U = iX$, with an operator norm distance between $U$
and $U_0$ of $\epsilon \simeq 8.5 \times 10^{-4}$. The braids $A_0$
and $B_0$ are the results of similar brute force searches to
approximate unitary operations $A$ and $B$ whose group commutator
satisfies $ABA^{-1}B^{-1}$ = $U U_0^{-1}$. The new braid $U_1 = A_0
B_0 A^{-1}_0 B^{-1}_0 U_0$ is then five times longer than $U_0$, and
the accuracy has improved so that the distance to the target gate is
now $\epsilon_1 \simeq 4.2 \times 10^{-5}$. Given the group
commutator structure of the $A_0 B_0 A_0^{-1} B_0^{-1}$ factor, the
winding of the $U_1$ braid is the same as the $U_0$ braid. Note that
when joining braids to form $U_1$ it is possible that elementary
braid operations from one braid will multiply their own inverses in
another braid, allowing the total braid to be shortened. Here we
have left these ``redundant" braids in $U_1$, as the careful reader
should be able to find.} \label{sk}
\end{figure}

What we have just described corresponds to one iteration of the
Solovay-Kitaev algorithm. Subsequent iterations are carried out
recursively. Thus, at the second level of approximation each
search within the $\epsilon$-net is replaced by the procedure
described above, and so on, so that at the $n^{th}$ level all
approximations are made at the $(n-1)^{st}$ level. The result of
this recursive process is a braid whose accuracy grows
superexponentially in $n$, with the distance to the desired gate
being of order $\epsilon_n \sim (c^2 \epsilon_0)^{(3/2)^n}$ at the
$n^{th}$ level of recursion, while the braid length grows only
exponentially in $n$, with $L \sim 5^n L_0$, where $L_0$ is a
typical braid length in the initial $\epsilon$-net. Thus, as the
distance of the approximate gate from the desired target gate,
$\epsilon$, goes to zero, the braid length grows only
polylogarithmically, with $L \sim \log^\alpha \frac{1}{\epsilon}$
where $\alpha = \ln 5 / \ln (3/2) \simeq 3.97$. While this scaling
is, of course, worse than the logarithmic scaling for brute force
searching, it is still only a polylogarithmic increase in braid
length which is sufficient for quantum computation. Similar
arguments\cite{harrow,dawson} can be used to show that the
classical computer time $t$ required to carry out the
Solovay-Kitaev algorithm also only scales polylogarithmically in
the desired accuracy, with $t \sim \log^\beta \frac{1}{\epsilon}$
where $\beta = \ln 3/\ln(3/2) \simeq 2.71$.

It is worth noting that there is a particularly nice feature of
this implementation of the Solovay-Kitaev algorithm when applied
to compiling three-braids. Recall that when carrying out two-qubit
gates it will be crucial to maintain the phase difference between
the total $q$-spin 1 and total $q$-spin 0 sectors of the
three-quasiparticle Hilbert space associated with a given
three-braid, and, according to (\ref{winding}), this can be done
by fixing the winding of the braid (modulo 10). Because of the
group commutator structure of the Solovay-Kitaev algorithm, the
winding of the $n^{th}$-level approximation $U_n$ will be the same
as that of the initial approximation $U_0$. This is because all
subsequent improvements involve multiplying this braid by group
commutators of the form $A_n B_n A_n^{-1} B_n^{-1}$ which
automatically have zero winding.  The phase relationship between
the total $q$-spin 1 and total $q$-spin 0 sectors is therefore
preserved at every level of the construction.

Fig.~\ref{sk} shows the application of one iteration of the
Solovay-Kitaev algorithm applied to finding a braid which
generates a unitary operation approximating $iX$. The braid
labeled $U_0$ is the result of a brute force search with $L=44$
corresponding to the best approximation shown in Fig.~\ref{brute}.
(Note that although this braid is drawn as a sequence of
elementary braid operations, it is topologically equivalent to a
weave. In fact precisely this braid, drawn explicitly as a weave,
is shown in Fig.~\ref{step2}.) The braids labeled $A_0$ and $B_0$
generate unitary operations which approximate operators $A$ and
$B$ whose group commutator gives $U U_0^{-1}$ where $U = iX$.
Finally, the braid labeled $U_1$ is the new, more accurate,
approximate weave.

\begin{figure}[t]
\includegraphics[scale=.15]{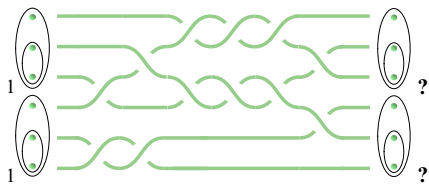}
\caption{(color online). Two encoded qubits and a generic braid.
Because quasiparticles are braided outside of their starting
qubits these braids will generally lead to leakage out of the
computational qubit space, i.e. the $q$-spin of each group of
three quasiparticles forming these qubits will in general no
longer be 1.} \label{twoqubits}
\end{figure}

\section{Two-qubit Gates}

We have seen that single-qubit gates are ``easy'' in the sense
that as long as we braid within an encoded qubit there will be no
leakage errors (the overall $q$-spin of the group of three
quasiparticles will remain 1). Furthermore, the space of unitary
operators acting on the three-quasiparticle Hilbert space
(essentially $SU(2)$) is small enough to find excellent
approximate braids by performing brute force searches and
subsequent improvement using the Solovay-Kitaev algorithm. We now
turn to the significantly harder problem of finding braids which
approximate entangling two-qubit gates.

\subsection{``Divide and Conquer" Approach}

Figure \ref{twoqubits} depicts six quasiparticles encoding two
qubits and a general braiding pattern. To entangle these qubits,
quasiparticles from one qubit must be braided around
quasiparticles from the other qubit, and this will inevitably lead
to leakage out of the encoded qubit space, (i.e. the overall
$q$-spin of the three quasiparticles constituting a qubit may no
longer be 1). Furthermore, the space of all operators acting on
the Hilbert space of six quasiparticles is much bigger than for
three, making brute force searching extremely difficult. Here the
unitary operations acting on this space are in $SU(5)\oplus
SU(8)$, (up to winding dependent phase factors as in
(\ref{winding})), which has 87 free parameters as opposed to 3 for
the three quasiparticle case of $SU(2)$.

Still, as a matter of principle, it is possible to perform a brute
force search of sufficient depth so that it corresponds to a fine
enough $\epsilon$-net to carry out the Solovay-Kitaev algorithm in
this larger space.\cite{kitaevbook}  This is essentially the program
outlined in Ref.~\onlinecite{freedman1} as an ``existence proof"
that universal quantum computation is possible; however, it is not
at all clear that, even if one could do this, it would be the most
efficient procedure for compiling braids. For the same amount of
classical computing power required to directly compile braids in
$SU(5)\oplus SU(8)$, we believe one can find much more efficient (in
the sense of having a more accurate computation with a shorter
braid) braids by breaking the problem into smaller problems, each
consisting of finding a specific three-braid embedded in the full
six-braid space. As we've shown above, these three-braids can then
be very efficiently compiled.

\begin{figure}[t]
\includegraphics[scale=.16]{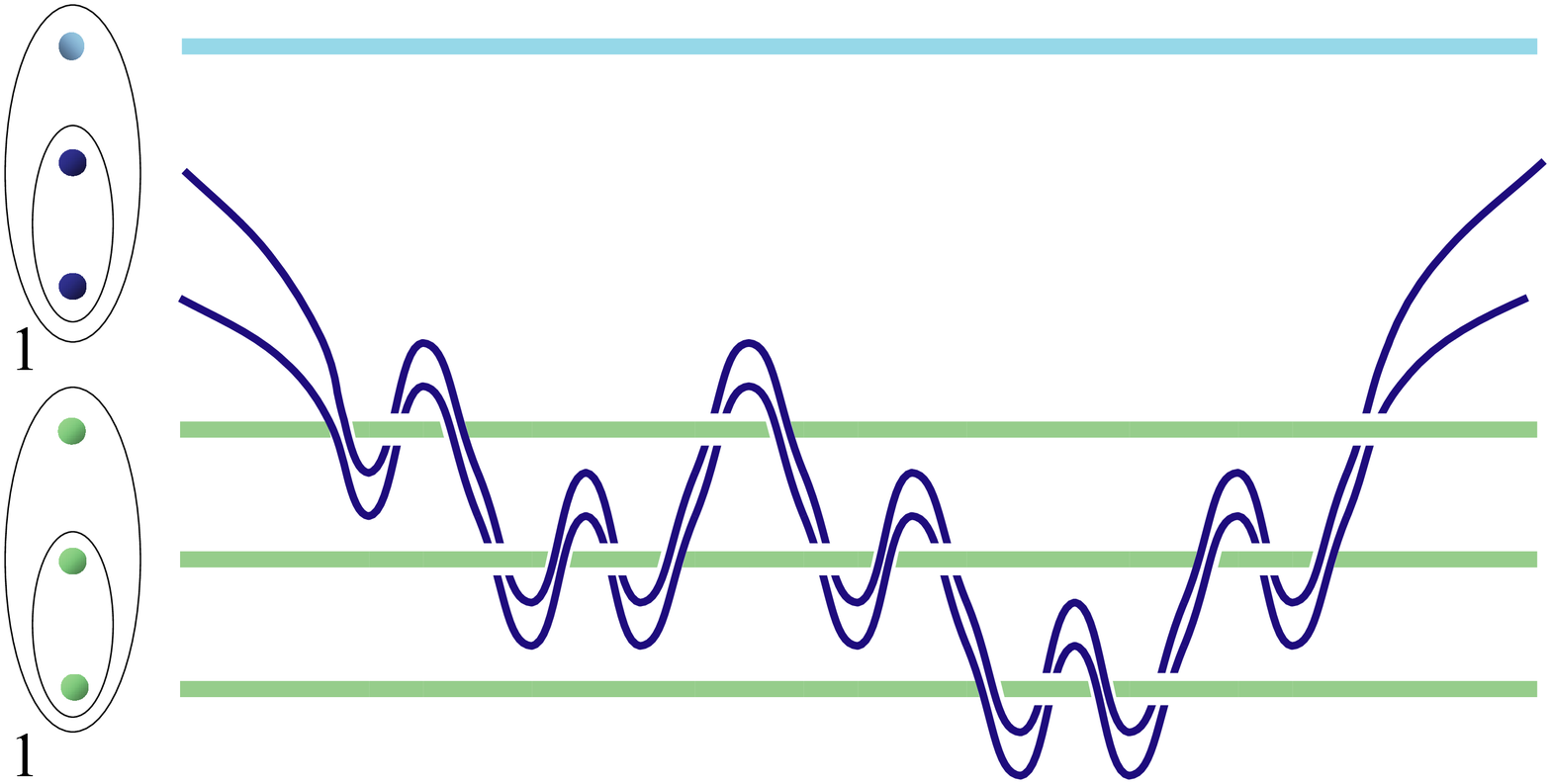}
\caption{(color online). A two-qubit gate construction in which a
pair of quasiparticles from the top (control) qubit is woven through
the bottom (target) qubit.  The mobile pair of quasiparticles is
referred to as the control pair and has a total $q$-spin of 0 if the
control qubit is in the state $|0_L\rangle$, and 1 if the control
qubit is in the state $|1_L\rangle$.  Since weaving an object with
total $q$-spin 0 yields the identity operation, this construction is
guaranteed to result in a transformation of the target qubit state
only if the control qubit is in the state $|1_L\rangle$.  Note that
in this and subsequent figures world-lines of mobile quasiparticles
will always be dark blue.} \label{controlpair}
\end{figure}

Here we present two classes of two-qubit gate constructions based
on this ``divide and conquer'' approach.  The first of these were
originally introduced by the authors in
Ref.~\onlinecite{bonesteel05} and are characterized by the weaving
of a {\it pair} of quasiparticles from one qubit through the
quasiparticles forming the second qubit.  The second class,
presented here for the first time, can be carried out by weaving
only a single quasiparticle from one qubit around one other
quasiparticle from the same qubit, and two quasiparticles from the
second qubit.

\subsection{Two-Quasiparticle Weave Construction}

We now review the two-qubit gate constructions first discussed in
Ref.~\onlinecite{bonesteel05}. The basic idea behind these
constructions is illustrated in Fig.~\ref{controlpair}.  This
figure shows two qubits and a braiding pattern in which a pair of
quasiparticles from the top qubit (the control qubit) is woven
through the quasiparticles forming the bottom qubit (the target
qubit). Throughout this braiding the pair is treated as a single
immutable object which, at the end of the braid, is returned to
its original position.

If, as in Fig.~\ref{controlpair}, we choose the pair of weft
quasiparticles to be the two quasiparticles whose total $q$-spin
determines the logical state of the qubit, then we refer to this
pair as the control pair.  We can then immediately see why this
construction naturally suggests itself.  If the control qubit is
in the state $|0_L\rangle$ the control pair will have total
$q$-spin 0, and weaving this pair through the target qubit will
have no effect. We are thus guaranteed that if the control qubit
is in the state $|0_L\rangle$ the identity operation is performed
on the target qubit.

\begin{figure}
\includegraphics[scale=.11]{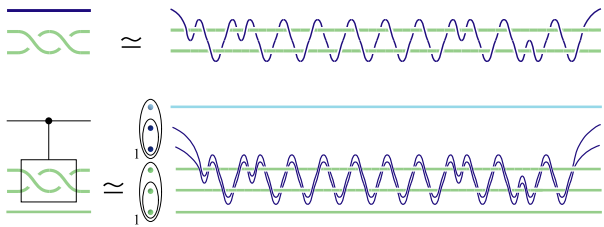}
\caption{(color online). An effective braiding weave, and a
two-qubit gate constructed using this weave.  The effective
braiding weave is a woven three-braid which produces a unitary
operation which is a distance $\epsilon \simeq 2.3 \times 10^{-3}$
from that produced by simply interchanging the two target
particles ($\sigma_1^2$). When the control pair is woven through
the target qubit using this weave the resulting two-qubit gate
approximates a controlled-$(\sigma_2^2)$ gate to a distance
$\epsilon \simeq 1.9 \times 10^{-3}$ or $\epsilon \simeq 1.6
\times 10^{-3}$ when the total $q$-spin of the two qubits is 0 or
1, respectively.} \label{effectivebraiding}
\end{figure}

The only non-trivial effect of this weaving pattern occurs when
the control qubit is in the state $|1_L\rangle$. In this case, the
control pair has total $q$-spin 1 and so behaves as a single
Fibonacci anyon.  The problem of constructing a two-qubit
controlled gate then corresponds to finding a weaving pattern in
which a single Fibonacci anyon weaves through the three
quasiparticles of the target qubit, inducing a transition on this
qubit without inducing leakage error out of the computational
qubit space, or at least keeping such leakage as small as required
for a particular computation. This reduces the problem of finding
a two-qubit gate to that of finding a weaving pattern in which one
Fibonacci anyon weaves around three others --- a problem involving
only four Fibonacci anyons.  However, following our ``divide and
conquer" philosophy, we will further narrow our focus to weaving a
single Fibonacci anyon through only two others at a time.

We define an ``effective braiding'' weave, to be a woven
three-braid in which the weft quasiparticle starts at the top
position, and returns to the top position at the end of the weave,
with the requirement that the unitary transformation it generates
be approximately equal to that produced by $m$ clockwise
interchanges of the two warp quasiparticles.  To find such weaves
we perform a brute force search, as outlined in Sec.~V, over
sequences $\{n_i\}$ which approximately satisfy
\begin{eqnarray}
\sigma_2\: U_{\rm weave}(\{n_i\})\: \sigma_2 \simeq \sigma_1^{m}.
\end{eqnarray}
If both sides of this equation are expressed using (\ref{winding})
it becomes evident that the winding of any effective braiding
weave must satisfy $W = m$ (modulo 10). Since the weft particle
starts and ends in the top position, $W$ must be even, thus
effective braiding weaves only exist for even $m$.

An example of an $m=2$ effective braiding weave found through a
brute force search is shown in Fig.~\ref{effectivebraiding}. The
corresponding unitary operation approximates that of interchanging
the two warp quasiparticles twice to a distance $\epsilon \sim
10^{-3}$. (This is a typical distance for a woven three-braid of
length $L \simeq 46$ which approximates a desired operation
--- precise distances of approximate weaves are given in the figure
captions.)  As for all approximate weaves considered here, the
Solovay-Kitaev algorithm outlined in Sec.~V.C can be used to
improve the accuracy of this weave so that $\epsilon$ can be made
as small as required with only a polylogarithmic increase in
length.

\begin{figure}[t]
\includegraphics[scale=.12]{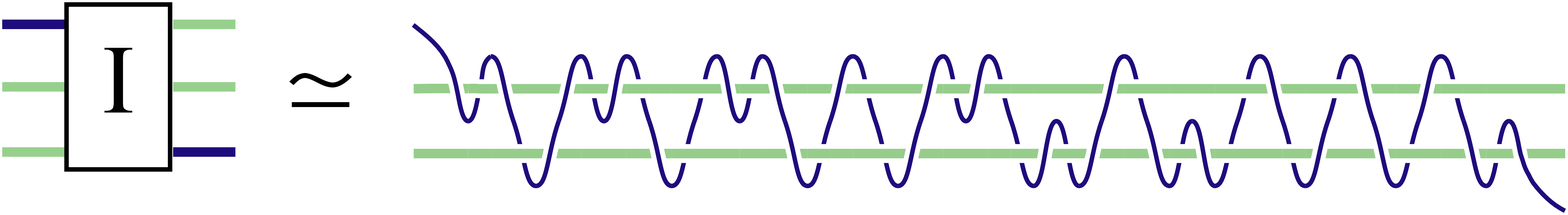}\\
\vskip .1in
\includegraphics[scale=.14]{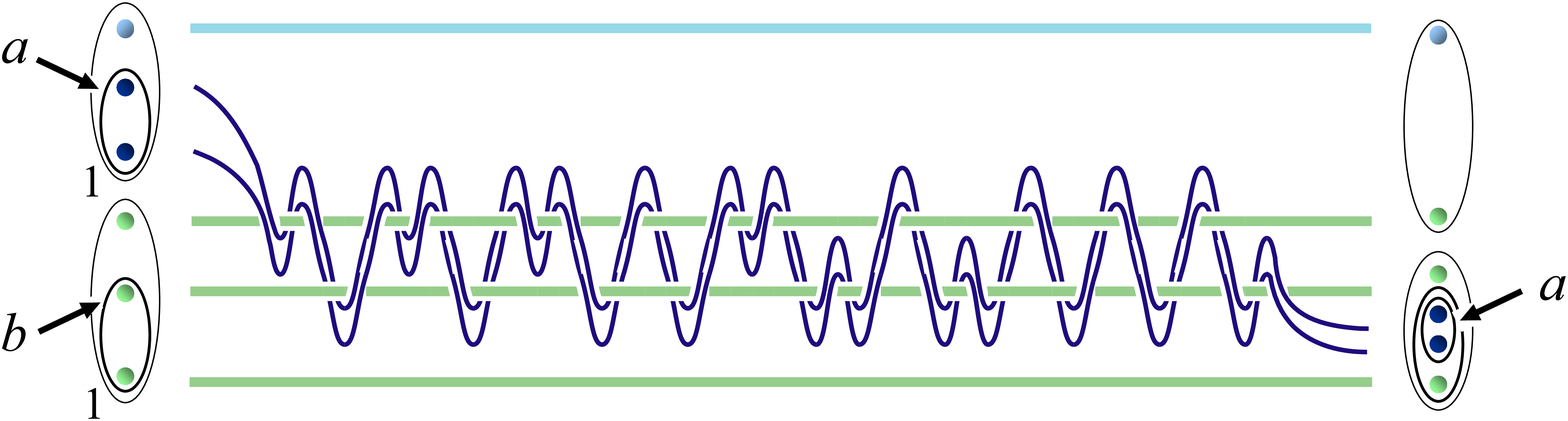}\\
\caption{(color online). An injection weave, and step one in our
injection based gate construction.  The box labeled $I$ represents
an ideal (infinite) injection weave which is approximated by the
weave shown to a distance $\epsilon \simeq 1.5 \times 10^{-3}$. In
step one of our gate construction, this injection weave is used to
weave the control pair into the target qubit. If the control qubit
is in the state $|1_L\rangle$ then $a=1$ and the result is to
produce a target qubit with the same quantum numbers as the
original, but with its middle quasiparticle replaced by the
control pair.} \label{step1}
\end{figure}

The construction of a two-qubit gate using this effective braiding
weave is also shown in Fig.~\ref{effectivebraiding}. In this
construction the control pair is woven through the top two
quasiparticles of the target qubit using this weave. As described
above, if the control qubit is in the state $|0_L\rangle$, the
control pair has $q$-spin 0 and the target qubit is unchanged.
But, if the control qubit is in the state $|1_L\rangle$, the
control pair has $q$-spin 1 and the action on the target qubit is
approximately equivalent to that of interchanging the top two
quasiparticles twice, with the approximation becoming more
accurate as the length of the effective braiding weave is
increased, either by deeper brute force searching or by applying
the Solovay-Kitaev algorithm. Because this effective braiding all
occurs within an encoded qubit, leakage errors can be reduced to
zero in the limit $\epsilon \rightarrow 0$. The resulting
two-qubit gate is then a controlled-$\sigma_2^2$ gate which
corresponds to controlled rotation of the target qubit through an
angle of $6 \pi/5$.

Unfortunately, due to the even $m$ constraint, it is impossible to
find an effective braiding gate which corresponds to a controlled
$\pi$ rotation of the target qubit. Such a gate would be equivalent
to a controlled-NOT gate up to single-qubit
rotations.\cite{nielsenbook} Nonetheless, it is known that any
entangling two-qubit gate, when combined with the ability to carry
out arbitrary single-qubit rotations, forms a universal set of
quantum gates.\cite{notcnot} Thus, the efficient compilation of
single-qubit operations described in Sec.~V and the effective
braiding construction just given provide direct procedures for
compiling any quantum algorithm into a braid to any desired
accuracy.

Although it can be used to form a universal set of gates, this
effective braiding construction is still rather restrictive. It is
clearly desirable to be able to directly compile a controlled-NOT
gate into a braid. We now give a construction which can be used to
efficiently compile any arbitrary controlled rotation of the target
qubit --- including a controlled-NOT gate. This construction is
based on a class of woven three-braids which we call ``injection
weaves".

\begin{figure}[t]
\includegraphics[scale=.12]{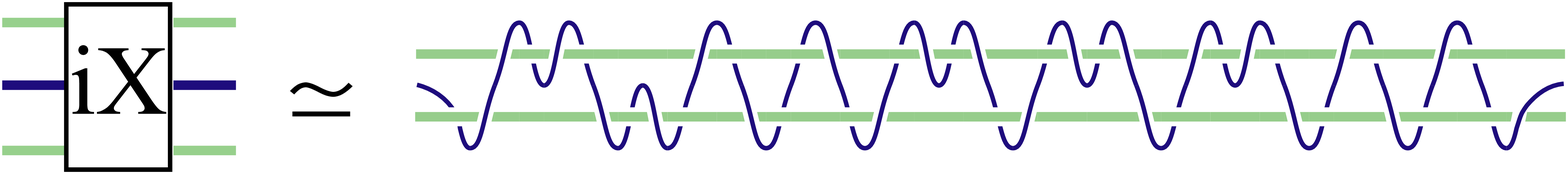}\\
\vskip .2in
\includegraphics[scale=.14]{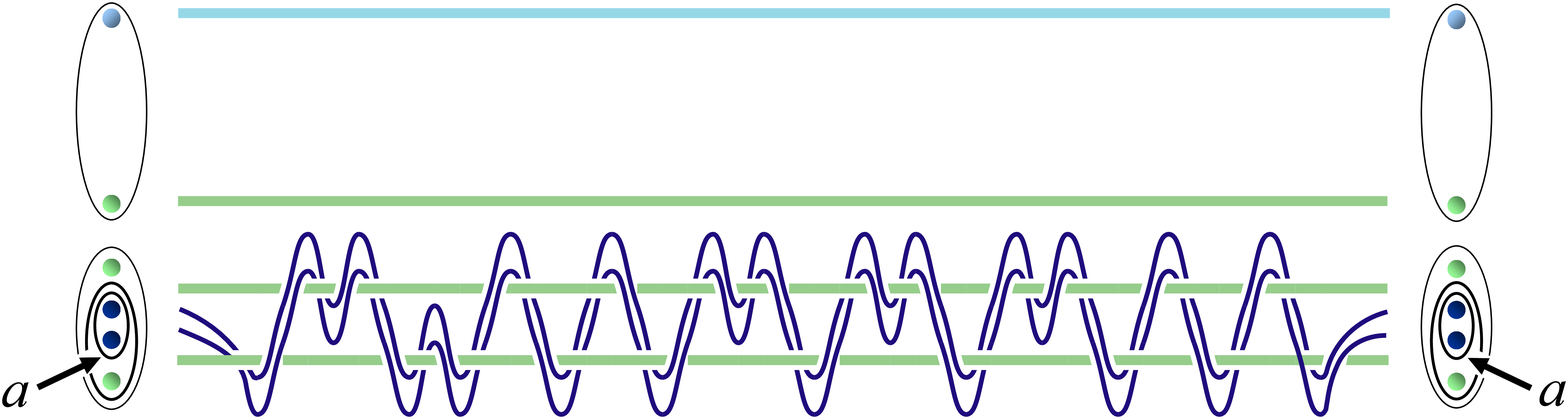}
\caption{(color online). A weave which approximates $iX$ (see
Eq.~\ref{ix}), and step two in our injection based construction.
The box labeled $iX$ represents an ideal (infinite) $iX$ weave
which is approximated by the weave shown to a distance $\epsilon =
8.5 \times 10^{-4}$ (this is the same weave which appears at the
top of Fig.~\ref{sk}). In step two of our gate construction the
control pair is woven within the injected target qubit, following
this weave, in order to carry out an approximate $iX$ gate when
$a=1$, as shown.} \label{step2}
\end{figure}

In an injection weave the weft quasiparticle again starts at the top
position but in this case ends at a different position.  At the same
time we require that the unitary operation generated by this weave
approximate the identity. Thus the effect of an injection weave is
to permute the quasiparticles involved without changing any of the
underlying $q$-spin quantum numbers of the system.

Comparing the identity matrix to (\ref{winding}) we see that any
three-braid approximating the identity must have winding $W = 0$
(modulo 10). The fact that this winding must be even implies that
the final position of the weft particle must be at the bottom of
the weave. Thus injection weaves correspond to sequences $\{n_i\}$
which approximately satisfy the equation,
\begin{eqnarray}
\sigma_1\: U_{\rm weave}(\{n_i\})\: \sigma_2  \simeq
\left(\begin{array}{cc|c} 1 & 0 &
\\ 0 & 1 &  \\ \hline &  & 1 \end{array}\right).
\end{eqnarray}
An injection weave obtained through brute force search is shown in
Fig.~\ref{step1}.  The unitary operation produced by this weave
approximates the identity operation to a distance $\epsilon \sim
10^{-3}$.

\begin{figure}[t]
\includegraphics[scale=.12]{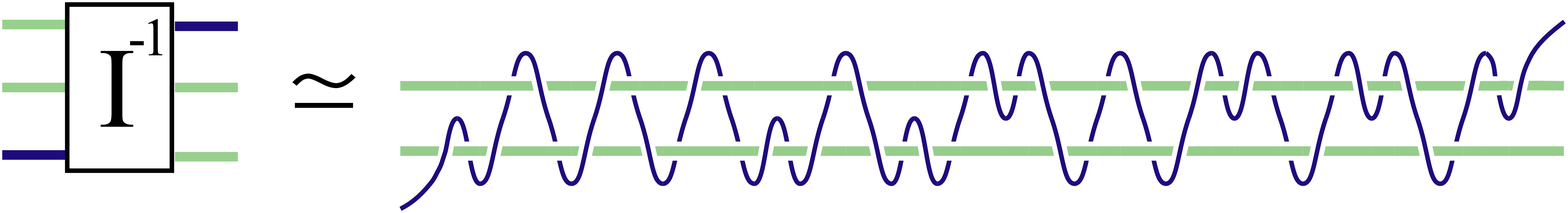}\\
\vskip .2in
\includegraphics[scale=.14]{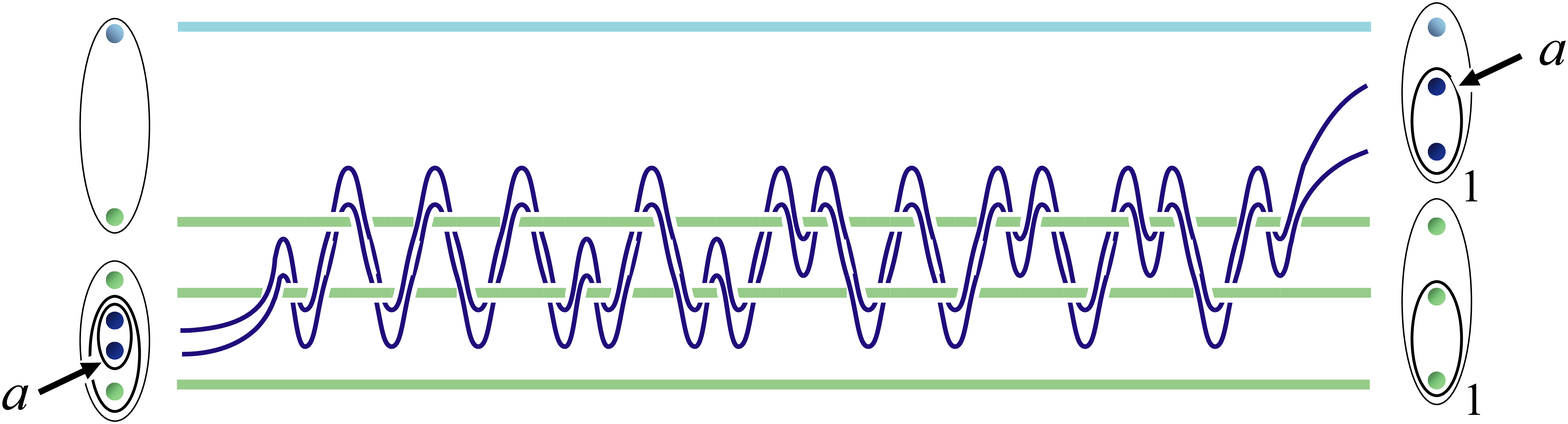}\\
\caption{(color online). An inverse injection weave and step three
in our injection based construction. The box labeled $I^{-1}$
represents an ideal (infinite) inverse injection weave which is
approximated by the the inverse of the injection weave shown in
Fig.~\ref{step1}, again to a distance $\epsilon \simeq 1.5 \times
10^{-3}$. This weave is used to extract the control pair out of
the injected target qubit and return it to the control qubit, as
shown.} \label{step3}
\end{figure}

\begin{figure*}
\centerline{\includegraphics[scale=1.1]{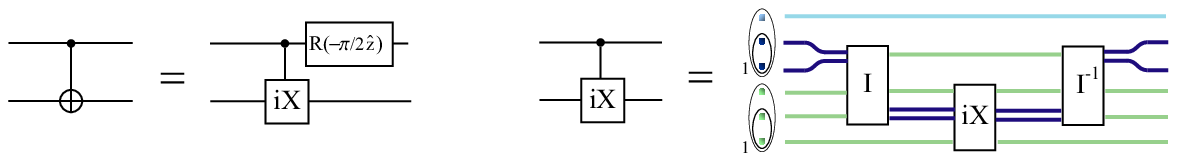}}
\centerline{\includegraphics[scale=1]{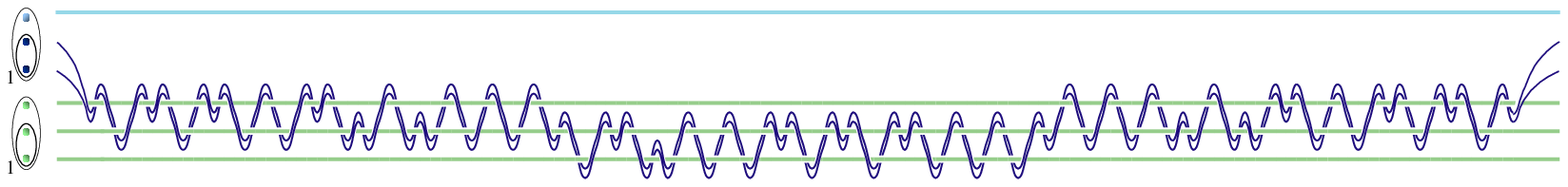}}
\caption{(color online). Injection-weave based compilation of a
controlled-NOT gate into a braid. A controlled-NOT gate can be
expressed as a controlled-$(iX)$ gate and a single-qubit operation
$R(-\pi/2\ \hat z) = \exp(i\pi\sigma_z/4)$ acting on the control
qubit. The single-qubit rotation can be compiled following the
procedure outlined in Sec.~V, and the controlled-$(iX)$ gate can
be decomposed into ideal injection ($I$), $iX$, and inverse
injection ($I^{-1}$) operations which can be similarly compiled.
The full approximate controlled-$(iX)$ braid obtained by replacing
$I$, $iX$ and $I^{-1}$ with the weaves shown in the previous three
figures is shown at bottom.  The resulting gate approximates a
controlled-$(iX)$ to a distance $\epsilon \simeq 1.8 \times
10^{-3}$ and $\epsilon \simeq 1.2 \times 10^{-3}$ when the total
$q$-spin of the two qubits is 0 or 1, respectively.} \label{fcnot}
\end{figure*}

Our two-qubit gate construction based on injection weaving is
carried out in three steps.  In the first step, also shown in
Fig.~\ref{step1}, the control pair is woven into the target qubit
using the injection weave.  If the control pair has total $q$-spin 1
(the only nontrivial case) the effect of this weave is merely to
replace the middle quasiparticle of the target qubit with the
control pair. Because the unitary operation approximated by the
injection weave is the identity, in the $\epsilon \rightarrow 0$
limit this injection is accomplished {\it without changing any of
the $q$-spin quantum numbers}.  The injected target qubit is
therefore (approximately) in the same quantum state as the original
target qubit.

In the second step of our construction, illustrated in
Fig.~\ref{step2}, we carry out an operation on the injected target
qubit by simply weaving the control pair within the target. Because
for $a=1$ all of this weaving takes place within the injected target
qubit, there will be no leakage error (again, strictly speaking,
only in the limit of an exact injection weave). The only constraint
on this weave is that the control pair must both start and end in
the middle position, and so it must have even winding.

If our goal is to produce a gate which is equivalent to a
controlled-NOT gate up to single-qubit rotations then we must apply
a $\pi$ rotation to the target qubit. Unfortunately, this cannot be
accomplished by any finite weave with even winding, so we must again
consider approximate weaves. Figure \ref{step2} shows the control
pair being woven through the injected target qubit using a weave
found by a brute force search which approximates a particular $\pi$
rotation --- the operator $i X$ defined in (\ref{ix}) --- to a
distance $\epsilon \sim 10^{-3}$ (this is, in fact, the same weave
shown at the top of Fig.~\ref{sk}).

The third step in our construction is the extraction of the
control pair from the target qubit.  This is accomplished, as
shown in Fig.~\ref{step3}, by applying the inverse of the
injection weave to the control pair.  The effect of this
extraction is to restore the control qubit to its original state,
and replace the control pair inside the target qubit with the
quasiparticle which originally occupied that position.

The full construction is summarized in Fig.~\ref{fcnot}, which
provides a recipe for compiling a controlled-NOT gate into a
two-quasiparticle weave.  A quantum circuit showing that a
controlled-NOT gate is equivalent to a controlled-$(iX)$ gate and a
single-qubit operation is shown in the top part of the figure. The
single-qubit operation can be compiled to whatever accuracy is
required following Sec.~V, and the controlled-$(iX)$ gate can be
decomposed into injection, $iX$, and inverse injection operations,
as is also shown in the top part of the figure.  These operations
can then all be similarly compiled following Sec.~V.

The full braid shown at the bottom of Fig.~\ref{fcnot} corresponds
to using the approximate woven three-braids shown in
Figs.~\ref{step1}-\ref{step3} to carry out a controlled-$(iX)$
gate.  In this braid, if the control qubit is in the state
$|0_L\rangle$ the control pair has total $q$-spin 0 and the
resulting unitary transformation is exactly the identity. However,
if the control qubit is in the state $|1_L\rangle$ the control
pair has total $q$-spin 1 and behaves like a single Fibonacci
anyon. This pair is then woven into the target qubit using an
injection weave, woven within the target in order to carry out the
$iX$ operation, and finally woven out of the target and back into
the control qubit using the inverse of the injection weave. The
resulting gate is therefore a controlled-$(iX)$ gate.

By replacing the $iX$ weave with an even winding weave which
carries out an arbitrary operation $U$ this construction will give
a controlled-$U$ gate.  The only restriction on $U$ is that its
overall phase must be consistent with (\ref{winding}) with even
winding $W$.  However, this phase can be easily set to any desired
value by applying the appropriate single-qubit rotation to the
control qubit, as in Fig.~\ref{fcnot}.

Finally, note that at no point in either the effective braiding or
injection weave constructions described above did we make reference
to the total $q$-spin of the two qubits involved. It follows that,
in the limit of exact effective braiding or injection weaves, the
action of the corresponding two-qubit gates on the computational
qubit space does not depend on the state of the external fusion
space associated with the $q$-spin 1 quantum numbers of each qubit
(see Sec.~IV). These gates will therefore not entangle the
computational qubit space with this external fusion space.

\subsection{One-Quasiparticle Weave Constructions}

We now show that two-qubit gates can be carried out with only a
single mobile quasiparticle.  This possibility follows from the
general result of Ref.~\onlinecite{simon06} that for any system of
nonabelian quasiparticles in which general braids are universal
for quantum computation (such as Fibonacci anyons), single
quasiparticle weaves are universal as well.  However, the ``proof
of principle" weaves constructed in that work were extremely
inefficient --- involving a huge number of excess operations. Here
we show how to efficiently construct a single-quasiparticle weave
corresponding to a controlled-NOT gate (up to single-qubit
rotations).

Our construction is based on a class of weaves which are similar
to injection weaves in that they can be used to swap two $q$-spin
1 objects --- where one object is a pair of Fibonacci anyons with
total $q$-spin 1 and the other object is a single Fibonacci anyon
--- while acting effectively as the identity operation so that
none of the other $q$-spin quantum numbers of the system are
disturbed. However, unlike injection weaves, this new class of
weaves accomplish this swap without moving the pair as a single
object, and in fact can be carried out by moving just one
quasiparticle.

The class of weaves we seek are those which approximate the
transformation
\begin{eqnarray}
U ((\bullet,\bullet)_a,\bullet)_c =
e^{i\phi}(\bullet,(\bullet,\bullet)_a)_c, \label{shift}
\end{eqnarray}
where $\phi$ is an overall (irrelevant) phase which does not
depend on $a$ or $c$.  The relevant case for showing the
similarity with injection is when $a=1$, for which the initial and
final states in (\ref{shift}) consist of two $q$-spin 1 objects
--- a single Fibonacci anyon and a pair of Fibonacci anyons with
total $q$-spin 1. If both these objects are represented as single
Fibonacci anyons then (\ref{shift}) can be written $U
(\bullet,\bullet)_c = e^{i\phi}(\bullet,\bullet)_c$. In this
representation $U$ therefore acts effectively as the identity
operation (times an irrelevant phase), similar to injection.

Using the $F$ matrix (\ref{fmatrix}) to expand the right hand side
of (\ref{shift}) in the $((\bullet,\bullet),\bullet)$ basis yields
\begin{eqnarray}
U ((\bullet,\bullet)_a,\bullet)_c = e^{i\phi} \sum_b F^{c}_{ab}
((\bullet,\bullet)_b,\bullet)_c. \label{uf}
\end{eqnarray}
Comparing this with the action of a unitary operation $U$ with
matrix representation
\begin{eqnarray}
U = \left(\begin{array}{cc|c} U^1_{00} & U^1_{01} & \\ U^1_{10} &
U^1_{11} & \\ \hline & & U^0_{11}\end{array}\right),
\end{eqnarray}
on the state $((\bullet,\bullet)_a,\bullet)_c$,
\begin{eqnarray}
U ((\bullet,\bullet)_a,\bullet)_c = \sum_b U^{c}_{ab}
((\bullet,\bullet)_b,\bullet)_c,
\end{eqnarray}
we see that the matrix representation of the $U$ we seek is
precisely the $F$ matrix (up to a phase): $U = e^{i\phi} F$. While
the $F$ matrix describes a ``passive" operation, i.e. a change of
basis, the operator $U$ can be viewed as an ``active" $F$ operation
which acts directly on the states of the Hilbert space. Note that,
since $F = F^{-1}$, we also have
\begin{eqnarray}
U (\bullet,(\bullet,\bullet)_a)_c =
e^{i\phi}((\bullet,\bullet)_a,\bullet)_c.\label{uinv}
\end{eqnarray}

\begin{figure}[t]
\includegraphics[scale=.12]{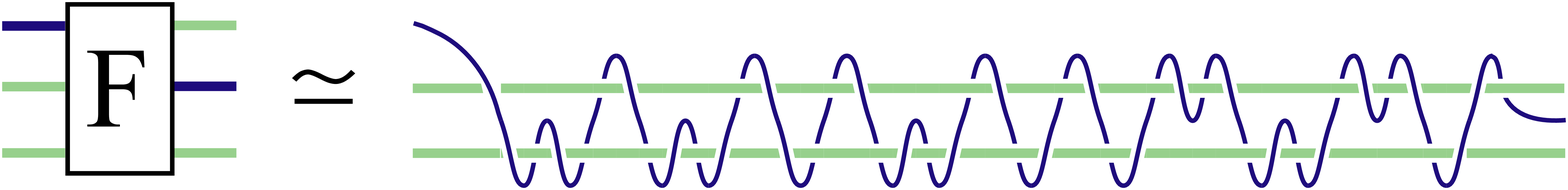}\\
\vskip .1in
\includegraphics[scale=.14]{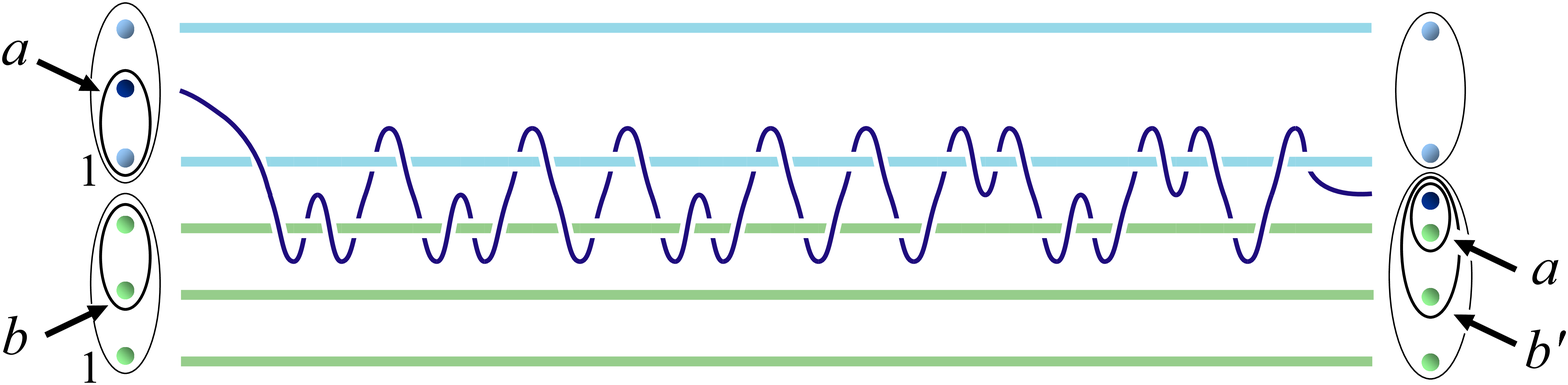}\\
\caption{(color online). An $F$ weave, and step one of our $F$ weave
based two-qubit gate construction.  The box labeled $F$ represents
an ideal (infinite) $F$ weave which is approximated by the weave
shown to a distance $\epsilon \simeq 3.1 \times 10^{-3}$. Applying
the $F$ weave to the initial two-qubit state, as shown, produces an
intermediate state with $q$-spins labeled $a$ and $b^\prime$ which
depend simply on $a$ and $b$ --- the initial states of the two
qubits (see Table I).}\label{fstep1}
\end{figure}

We will refer to weaves which approximate the operation
(\ref{shift}) (and thus also (\ref{uinv})) as $F$ weaves.   As we
have seen, the unitary operation $U$ produced by an $F$ weave need
only approximate the $F$ matrix (\ref{fmatrix}) up to an overall
irrelevant phase. To be consistent with (\ref{winding}) this phase
must be $-1$, as can be seen by writing the matrix $-F$ as
\be - F = \left(\begin{array}{c|c} \pm i \left(\begin{array}{cc}
 \pm i\tau & \pm i\sqrt{\tau}\\ \pm i\sqrt{\tau}& \mp i\tau
 \end{array}\right) \\ \hline &  -1 \end{array}\right),
\label{fweavematrix}\ee
where a factor of $\pm i$ has been pulled out of the upper left
2$\times$2 block, leaving an $SU(2)$ matrix ($\det = \tau^2 + \tau
= 1$). Comparing (\ref{fweavematrix}) with (\ref{winding}), it is
also evident that any $F$ weave must have winding $W =$ 5 (modulo
10), which is necessarily odd.

The fact that $F$ weaves must have an odd number of windings implies
that if the weft quasiparticle starts at the top position of the
weave it must end at the middle position.  For this choice the $F$
weave must then approximately satisfy the equation
\begin{eqnarray}
U_{\rm weave}(\{n_i\})\: \sigma_2 \simeq - F.
\end{eqnarray}
The result of a brute force search for an $F$ weave which
approximates the operation $-F$ to a distance $\epsilon \sim
10^{-3}$ is shown in Fig.~\ref{fstep1}.

The first step in our single-quasparticle weave construction is
the application of an $F$ weave to two qubits, also shown in
Fig.~\ref{fstep1}. Note that in this figure for convenience we
have made a change of basis on the bottom qubit, so that the pair
which determines its state (the control pair) consists of the top
two quasiparticles within it rather than the bottom two. There is
no loss of generality in doing so since this just corresponds to a
single-qubit rotation on the bottom qubit.

With this basis choice the initial state of the two qubits is
determined by the $q$-spins of their respective control pairs which
are indicated in Fig.~\ref{fstep1} as $a$ (top qubit) and $b$
(bottom qubit). After carrying out the $F$ weave, taking the middle
quasiparticle of the top qubit as the weft quasiparticle and weaving
it around both the bottom quasiparticle of the top qubit and the top
quasiparticle of the bottom qubit, the resulting state (again,
strictly speaking, only in the limit of an exact $F$ weave) is shown
at the end of the two-qubit weave in Fig.~\ref{fstep1}. From
(\ref{uinv}) it follows that the newly positioned weft quasiparticle
and the quasiparticle beneath will have total $q$-spin $a$. When the
quasiparticle beneath these two is also included, the three
quasiparticles form what we will refer to as the intermediate state,
$(\bullet,(\bullet,\bullet)_a)_{b^\prime}$, where the total $q$-spin
of all three quasiparticles, $b^\prime$, has a well-defined value
provided $a$ and $b$ are well defined, as we now show.

\begin{table}[b]
\begin{tabular}{c|c|c|c|c}
 ~~~~~$a$~~~~~ & ~~~~~$b$~~~~~ & &~~~~~$b^\prime$~~~~~ & Phase
 Factor\\
\tableline
0 & 0  & $b^\prime = 1$ & 1 & $e^{i\alpha}$ \\
0 & 1  &                & 1 & $e^{i\alpha}$ \\  \tableline
1 & 0  & $b^\prime = b$ & 0 & 1\\
1 & 1  &                & 1 & $e^{-i\alpha}$
\end{tabular}
\caption{Values of $b^\prime$ for different values of $a$ and $b$
after applying the $F$ weave as shown in Fig.~\ref{fstep1}, and the
phase applied to the resulting state by a phase weave with zero
winding. The value of $b^\prime$ is determined by the fact that
$b^\prime = 1$ when $a=0$ and $b^\prime = b$ when $a=1$, as shown in
the text.}\label{table}
\end{table}

\begin{figure}[t]
\includegraphics[scale=.12]{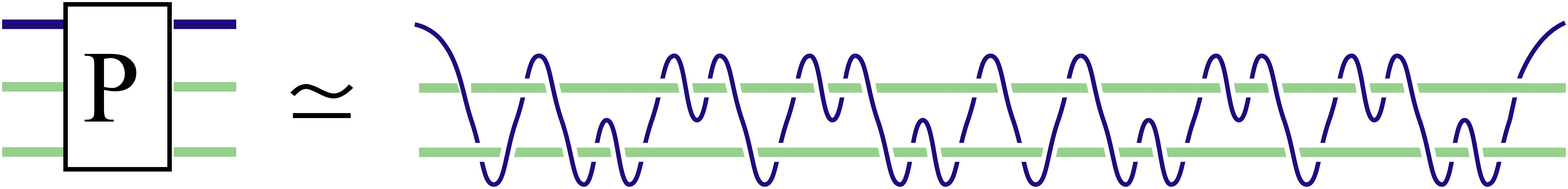}\\
\vskip .2in
\includegraphics[scale=.14] {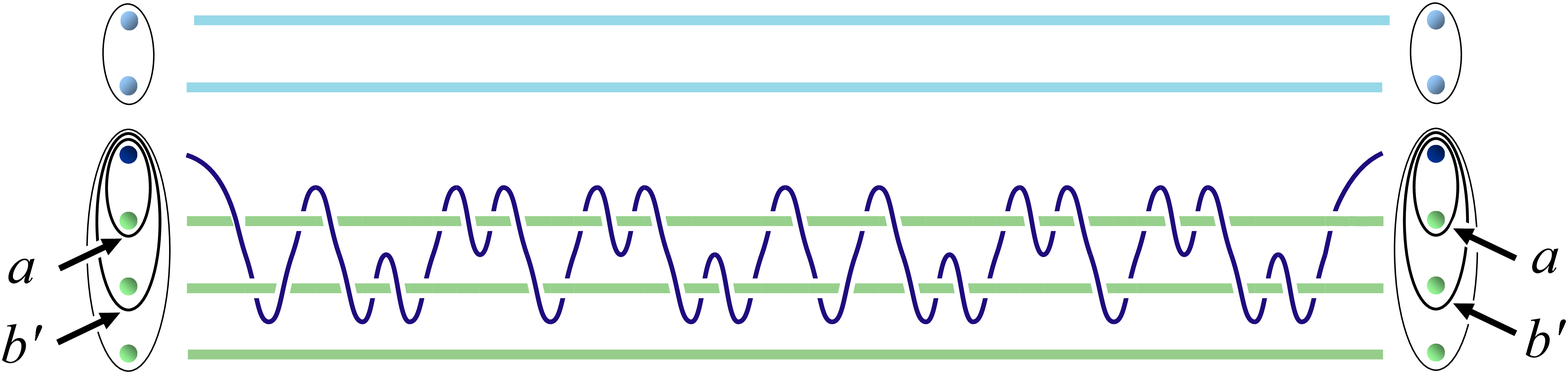}\\
\caption{(color online). A phase weave with $\alpha = \pi$ (see
text) which gives a $\pi$ phase shift to the intermediate state
when $b^\prime = 1$, and step two of our $F$ weave based
construction. The box labeled $P$ represents an ideal (infinite)
$\alpha = \pi$ phase weave which is approximated by the weave
shown to a distance $\epsilon \simeq 1.9 \times 10^{-3}$. Applying
this phase weave to the intermediate state created by the $F$
weave, as shown, results in a $b^\prime$ dependent $\pi$ phase
shift (see Table I with $\alpha = \pi$).} \label{fstep2}
\end{figure}

First consider the case $a=1$.  As described above, the effect of
the $F$ weave is then similar to that of the injection weave from
the previous construction --- it replaces the topmost quasiparticle
in the bottom qubit with a pair of quasiparticles with $q$-spin 1,
and the bottommost pair of quasiparticles in the top qubit (which
also has total $q$-spin 1) with a single quasiparticle, without
changing any of the other $q$-spin quantum numbers of the system. In
the limit of an ideal $F$ weave, this means that the $b$ quantum
number does not change after this swap and so $b^\prime = b$. The
case $a=0$ is simpler, since in this case the intermediate state is
$(\bullet,(\bullet,\bullet)_0)_{b^\prime}$ for which the fusion
rules (\ref{fibfusion}) imply $b^\prime = 1$, regardless of the
value of $b$. The resulting dependence of $b^\prime$ on $a$ and $b$
is summarized in Table \ref{table}.

\begin{figure}[t]
\includegraphics[scale=.12]{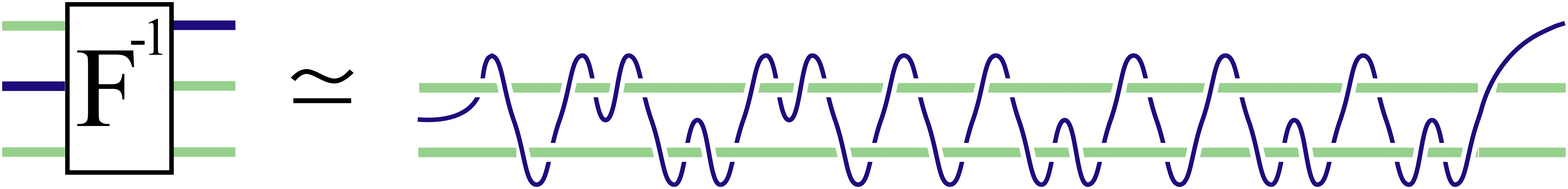}\\
\vskip .2in
~~~~\includegraphics[scale=.14] {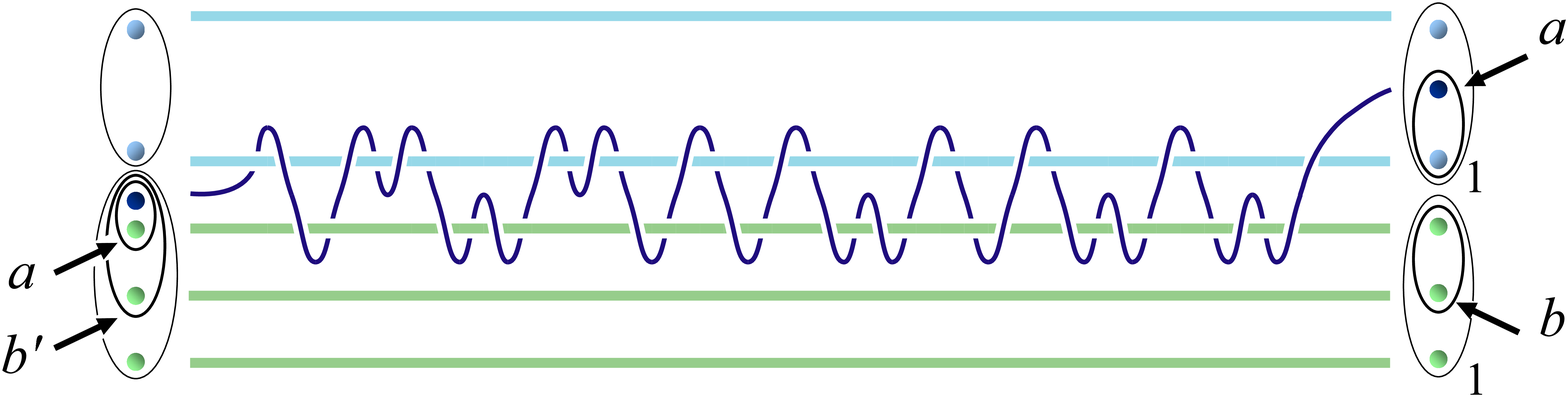}\\
\caption{(color online). An inverse $F$ weave and step three in
our $F$ weave construction. The box labeled $F^{-1}$ is an ideal
(infinite) inverse $F$ weave which is approximated by the inverse
of the $F$ weave shown in Fig.~\ref{fstep1}, again to a distance
$\epsilon \simeq 3.1 \times 10^{-3}$. By applying the inverse $F$
weave to the state obtained after applying the phase weave, as
shown, the two qubits are returned to their initial states, but
now with an $a$ and $b$ dependent phase factor (see Table I).}
\label{fstep3}
\end{figure}

Having used the $F$ weave to create the intermediate state
$(\bullet,(\bullet,\bullet)_a)_{b^\prime}$, the next step in our
construction is the application of a weave which performs an
operation on this state which does not change $a$ and $b^\prime$ but
which does yield an $a$ and $b^\prime$ dependent phase factor. After
carrying out such a weave, which we will refer to as a phase weave,
we can then apply the inverse of the $F$ weave to restore the two
qubits to their initial states $a$ and $b$.

\begin{figure*}
\centerline{\includegraphics[scale=1.1]{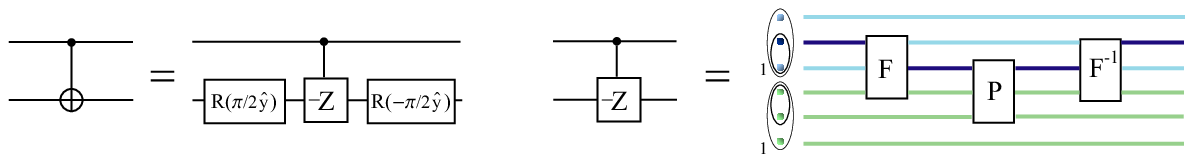}}
\centerline{\includegraphics[scale=1]{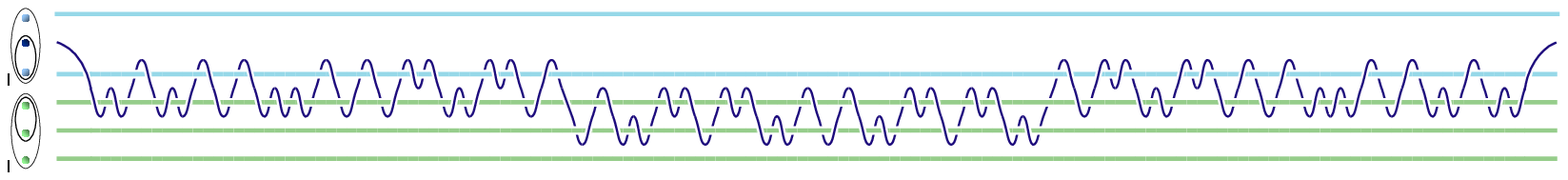}}
\caption{(color online). $F$ weave based compilation of a
controlled-NOT gate into a braid. A controlled-NOT gate is
equivalent to a controlled-$(-Z)$ gate with the single-qubit
operation $R(\pi/2\ \hat y) = \exp(-i\pi \sigma_y/4)$ and its
inverse applied to the target qubit before and after the
controlled-$(-Z)$.  Again, the single-qubit operations can be
trivially compiled, and the controlled-$(-Z)$ gate decomposed into
ideal $F$, phase $(P)$, and inverse $F$ ($F^{-1}$) weaves which
can be similarly compiled. The full approximate controlled-$(-Z)$
weave obtained by replacing $F$, $P$ and $F^{-1}$ with the
approximate weaves shown in the previous three figures is shown at
bottom.  The resulting gate approximates a controlled-$(-Z)$ to a
distance $\epsilon \simeq 4.9 \times 10^{-3}$ and $\epsilon \simeq
3.2 \times 10^{-3}$ when the total $q$-spin of the two qubits is 0
or 1, respectively.} \label{ffcnot}
\end{figure*}

For any phase weave we will require that the weft quasiparticle
both start and end in the top position so that when we join it to
the $F$ weave and its inverse there will be a single weft
quasiparticle throughout the entire gate construction. The phase
weave must therefore have even winding, and with no loss of
generality we can consider the case for which the winding
satisfies $W = 0$ (modulo 10). The unitary operation produced by
such a phase weave must then approximately satisfy the equation
\begin{eqnarray}
\sigma_2 U_{\rm weave}(\{n_i\})\sigma_2 \simeq F
\left(\begin{array}{cc|c} e^{i\alpha} & 0 & \\ 0 & e^{-i\alpha} &
\\ \hline & & 1
\end{array}\right) F^{-1},
\end{eqnarray}
where the $F$ matrices are needed to change the Hilbert space
basis from that in which the operation produced by the phase braid
must be diagonal, (the $(\bullet,(\bullet,\bullet))$ basis), to
that in which the $\sigma_1$ and $\sigma_2$ matrices are defined,
(the $((\bullet,\bullet),\bullet)$ basis).

We will see that a phase weave with $\alpha = \pi$ produces a
two-qubit gate which is equivalent to a controlled-NOT gate up to
single-qubit rotations. The result of a brute force search for
such a phase weave which approximates the desired operation to a
distance $\epsilon \sim 10^{-3}$ is shown in Fig.~\ref{fstep2}.
This figure also shows the action of the phase weave on the
intermediate state produced in Fig.~\ref{fstep1}.  In this weave,
the weft quasiparticle is now woven through the two quasiparticles
beneath it, and returns to its original position. Because the
phase weave produces a diagonal operation in the basis shown for
the intermediate state, it does not change the values of $a$ and
$b^\prime$.  Its only effect is to give a phase factor of
$e^{i\alpha}$ to the state with $a = 0$ (which necessarily has
$b^\prime = 1$) and $e^{-i\alpha}$ to the state with $a=1$ and
$b^\prime = 1$.  The state with $a=1$ and $b^\prime = 0$ is
unchanged.  These phase factors are also shown in Table
\ref{table}.

The final step in this construction is to perform the inverse of
the $F$ weave to return the two qubits to their original states.
This is shown in Fig.~\ref{fstep3}.  In the limit of exact $F$ and
phase weaves, the resulting operation on the computational qubit
space in the basis $ab = \{00, 01, 10, 11\}$ is then,
\be
U = \left(
\begin{array}{cccc}
e^{i\alpha} & 0 & 0 & 0 \\
0 & e^{i\alpha} & 0 & 0 \\
0 & 0 & 1 & 0 \\
0 & 0 & 0 & e^{-i\alpha}
\end{array}
\right). \label{fgate} \ee
If we take the top qubit to be the control qubit, and the bottom
qubit to be the target qubit, then this gate corresponds, up to an
irrelevant overall phase, to a controlled-($e^{-i 3 \alpha/2 }
e^{i \alpha \sigma_z/2})$ operation. For the case $\alpha = \pi$
this is a controlled-$(-Z)$ gate (where $Z = \sigma_z$), i.e. a
controlled-Phase gate, which, up to single-qubit rotations, is
equivalent to a controlled-NOT gate.

The full $F$ weave based gate construction is summarized in
Fig.~\ref{ffcnot}. A quantum circuit showing a controlled-NOT gate
in terms of a controlled-$(-Z)$ gate and two single-qubit operations
is shown in the top part of the figure. As in our injection based
construction, the single-qubit operations can be compiled to
whatever accuracy is required following the procedure outlined in
Sec.~V. The controlled-$(-Z)$ gate can then be decomposed into ideal
$F$, phase, and inverse $F$ weaves as is also shown in the top part
of the figure. Woven three-braids which approximate these operations
can then be compiled to whatever accuracy is required, again
following Sec.~V. The full controlled-$(-Z)$ weave corresponding to
using the approximate $F$ and phase weaves shown in
Figs.~\ref{fstep1}-\ref{fstep3} is shown in the bottom part of the
figure.

Finally, in this construction, as for the constructions described
in Sec.~VI.B, we at no point made reference to the total $q$-spin
of the two qubits involved. Thus, in the limit of exact $F$ and
phase weaves, the action of the two-qubit gates constructed here
will not entangle the computational qubit space with the external
fusion space associated with the $q$-spin 1 quantum numbers of
each qubit.

\section{What's special about $k=3$?}

All of the gate constructions discussed in this paper exploit the
fact that the braiding and fusion properties of a pair of Fibonacci
anyons are either trivial if their total $q$-spin is 0, or
equivalent to those of a single Fibonacci anyon if their total
$q$-spin is 1. The fact that these are the only two possibilities is
a special property of the Fibonacci anyon model, and hence also the
$SU(2)_3$ model, given their effective equivalence. It is then
natural to ask to what extent our constructions can be generalized
to $SU(2)_k$ CSW theories for different values of the level
parameter $k$.

Of course we know from the results of Freedman et
al.\cite{freedman2} that the $SU(2)_k$ representations of the
braid group are dense for $k=3$ and $k>4$.  Thus, for example,
braids which approximate controlled-NOT gates on encoded qubits
exist and can, in principle, be found for all these $k$ values.
However, we will show below that things are somewhat simpler for
the case $k=3$. Specifically we will show that for $k=3$, and only
$k=3$, it is possible to carry out two-qubit entangling gates by
braiding only four quasiparticles, as, for example, in our
effective braiding and $F$ weave constructions.

\begin{figure}[t]
\includegraphics[scale=.2]{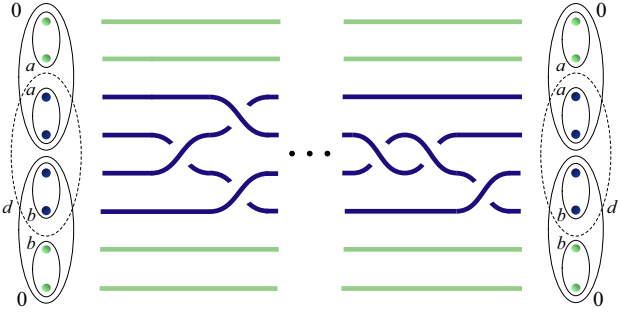}
\caption{(color online). Two four-quasiparticle qubits and a
braiding pattern in which only two quasiparticles from each qubit
are braided. Here the quasiparticles are $SU(2)_k$ excitations with
$q$-spin 1/2. The state of the top qubit is determined by the total
$q$-spin of the quasiparticle pairs labeled $a$ and the state of the
bottom qubit is determined by the total $q$-spin of the
quasiparticle pairs labeled $b$. The overall $q$-spin of the four
braided quasiparticles is $d$, (a dashed oval is used because when
$a=b=1$ these quasiparticles will not be in a $q$-spin eigenstate).
For this braid to produce no leakage errors, the unitary operation
it generates must be diagonal in $a$ and $b$, though it can, of
course, result in an $a$ and $b$ dependent phase factor. For $k >
3$, $d$ can take the values 0, 1 or 2, while for $k=3$ the only
allowed values for $d$ are 0 and 1. The existence of the $d=2$ state
for $k > 3$ makes it impossible to carry out an entangling two-qubit
gate by braiding only four quasiparticles (see text).} \label{proof}
\end{figure}

Consider a pair of $SU(2)_k$ four-quasiparticle qubits as shown in
Fig.~\ref{proof}.  Here each quasiparticle is assumed to have
$q$-spin 1/2 and the total $q$-spin of each qubit is required to
be 0. The state of a given qubit is then determined by the
$q$-spin of either the topmost or bottommost pair of
quasiparticles within it, where, from the $SU(2)_k$ fusion rules
(\ref{fusion}), the $q$-spin of each pair must be the same for the
total $q$-spin of the qubit to be 0. Thus, in Fig.~\ref{proof},
the state of the top qubit is determined by the $q$-spin labeled
$a$ and the state of the bottom qubit is determined by the
$q$-spin labeled $b$, where, again from the fusion rules
(\ref{fusion}), $a$ and $b$ can be either 0 or 1.

If we are only allowed to braid the middle four quasiparticles, as
shown in Fig.~\ref{proof}, then the total $q$-spin of the two
topmost quasiparticles of the top qubit and the two bottommost
quasiparticles of the bottom qubit will remain, respectively, $a$
and $b$. It follows that if the two qubits are to remain in their
computational qubit spaces, the total $q$-spin of the two topmost
and two bottommost quasiparticles that {\it are} being braided must
also remain, respectively, $a$ and $b$.  (If this were not the case,
the fusion rules (\ref{fusion}) would imply that the total $q$-spin
of the four quasiparticles forming each qubit would no longer be 0).
Thus, in order for there to be no leakage errors after braiding
these four quasiparticles, the resulting operation must be diagonal
in $a$ and $b$.

It is important to note that this result, and the results that
follow, hold not just for four-quasiparticle qubits, but also for
$SU(2)_k$ versions of the three-quasiparticle qubits used
throughout this paper. This is because, as pointed out in Sec.~IV,
any gate acting on a pair of three-quasiparticle qubits must
result in an operation on the computational qubit space which is
independent of the state of the external fusion space associated
with the fact that each qubit has total $q$-spin 1/2, (here the
total $q$-spin of a three-quasiparticle qubit is 1/2 rather than 1
because we are using $SU(2)_k$ quantum numbers and assuming each
quasiparticle has $q$-spin 1/2 --- see Fig.~\ref{notation}(b)). It
is therefore sufficient to consider the special case when the
state of two three-quasiparticle qubits corresponds to that of the
two four-quasiparticle qubits shown in Fig.~\ref{proof}, but with
the topmost and bottommost quasiparticles removed. The above
arguments then imply any leakage free operation produced by
braiding the four middle quasiparticles must be diagonal in $a$
and $b$.

Now consider the four middle quasiparticles we are allowed to braid.
A basis for the Hilbert space of these quasiparticles can be taken
to be one labeled by the $q$-spin quantum numbers $a$ and $b$, as
well as the total $q$-spin of all four quasiparticles which we
denote $d$ (see Fig.~\ref{proof}). For $k>3$ the fusion rules
(\ref{fusion}) imply this total $q$-spin $d$ can be equal to 0, 1 or
2, while for $k=3$ it can only be equal to 0 or 1. We will see that
this truncation of the $d=2$ state is the crucial property of the
$k=3$ theory which makes our $F$ weave and effective braiding
constructions possible.

It is convenient at this stage to restrict ourselves to braids
with zero total winding (i.e. equal numbers of clockwise and
counterclockwise exchanges).  For such braids, arguments similar
to those used to derive (\ref{winding}) can be used to show the
unitary operation enacted on the $d=0, 1$ and 2 sectors must each
have determinant 1. There is no loss of generality in restricting
ourselves to such braids, since a braid with arbitrary winding can
always be turned into one with zero winding by adding the
appropriate number of interchanges to either the two topmost or
two bottommost of the braiding quasiparticles at either the
beginning or end of the braid. These added interchanges will all
be within encoded qubits and so correspond to single-qubit
rotations which will not produce any entanglement between the two
qubits.

If we restrict ourselves to braids with zero winding and insist
that these braids approximate gates with zero leakage error
--- which, as shown above, implies the gate must be diagonal in
the $a$ and $b$ quantum numbers --- then in the $abd = \{000, 110,
011,101, 111, 112\}$ basis the unitary transformation acting on the
Hilbert space of the four braiding quasiparticles must have the form
\begin{eqnarray}
U = \left(\begin{array}{cc|ccc|c}
e^{i\alpha} & 0 &  &  &  &  \\
0 & e^{-i\alpha} &  &  &  &  \\
\hline
 &  & e^{i\beta} & 0 & 0 & \\
 &  & 0 & e^{i\gamma} & 0 &  \\
 &  & 0 & 0 & e^{-i(\beta+\gamma)} & \\
\hline  &  &  &  &  & 1
\end{array} \right),
\label{kgt3}
\end{eqnarray}
where we have required that the $d=0$, 1 and 2 blocks all have
determinant 1, (in particular, the $d=2$ block is simply 1).

Note that the case $a=b=1$ has three entries in this matrix,
corresponding to the three possible values for the total $q$-spin
quantum number $d$.  For this gate to produce no leakage error,
the phase factors in all three of these sectors must be the same.
To see this note that one can expand the relevant
eight-quasiparticle state in terms of basis states with
well-defined values of $d$ as follows
\begin{eqnarray}
&&(((\bullet,\bullet)_1,(\bullet,\bullet)_1)_0,
((\bullet,\bullet)_1,(\bullet,\bullet)_1)_0)_0 \nonumber\\
&&~= \sum_{d=0}^2 F_d\ \left((\bullet,\bullet)_1,
\left((\bullet,\bullet)_1, (\bullet,\bullet)_1\right)_d,
(\bullet,\bullet)_1\right)_0,
\end{eqnarray}
where standard quantum group methods\cite{slingerland01,fuchsbook}
can be used to compute the coefficients $F_d$, with the result
\begin{eqnarray}
F_0 = \frac{1}{[3]_q},\ \ F_1 = \frac{\sqrt{[3]_q}}{[3]_q},\ \ F_2 =
\frac{\sqrt{[5]_q}}{[3]_q}.
\end{eqnarray}
Here we have introduced the $q$-integers $[m]_q \equiv (q^{m/2}
-q^{-m/2})/(q^{1/2}-q^{-1/2})$, where $q = e^{i2\pi/(k+2)}$ is the
deformation parameter.

For $k > 3$ all three $F_d$ coefficients are nonzero.  Thus, in
order for the action of (\ref{kgt3}) on the $a=b=1$ state to
produce the same state back (up to a phase), the projection of
this state in the three $d$ sectors must all acquire the same
phase. This implies that $\alpha = 0$ and  $\beta = - \gamma$. The
resulting unitary operation must therefore take the form
\begin{eqnarray}
U =  \left(\begin{array}{cc|ccc|c}
1 & 0 &  &  &  &  \\
0 & 1 &  &  &  &  \\
\hline
 &  & e^{i\beta} & 0 & 0 &  \\
 &  & 0 & e^{-i\beta} & 0 &  \\
 &  & 0 & 0 & 1 & \\
\hline  &  &  &  &  & 1
\end{array} \right),
\end{eqnarray}
which corresponds to the following two-qubit gate in the $ab = \{00,
01, 10, 11\}$ basis,
\begin{eqnarray}
U^{gate}_{k>3} = \left(\begin{array}{cccc} 1 & 0 & 0 & 0 \\
0 & e^{i\beta} & 0 & 0 \\
 0 & 0 & e^{-i\beta} & 0 \\
0 & 0 & 0 & 1
\end{array}\right).\label{ukgt3}
\end{eqnarray}
This gate is simply the tensor product of two single-qubit
rotations, $U^{gate}_{k>3} = e^{-i\beta \sigma_z^{(1)}/2} \otimes
e^{i\beta \sigma_z^{(2)}/2}$.  Thus we see that for $k>3$ any
two-qubit gate constructed by braiding only four quasiparticles
for which there is no leakage error must necessarily also produce
no entanglement.

For $k=3$ this argument breaks down because the $d=2$ sector of
the braiding quasiparticles is not present.  In this case,
following the same argument as above, in the $abd = \{000, 110,
011, 101, 111\}$ basis the allowed leakage free unitary
transformations which can be produced by braiding the four middle
quasiparticles must be of the form (again taking the case of zero
winding),
\begin{eqnarray}
U = \left(\begin{array}{cc|ccc}
e^{i\alpha} & 0 &  &  &  \\
0 & e^{-i\alpha} &  &  &  \\
\hline
 &  & e^{i\beta} & 0 & 0  \\
 &  & 0 & e^{i(\alpha-\beta)} & 0  \\
 &  & 0 & 0 & e^{-i\alpha} \\
\end{array} \right),
\end{eqnarray}
which corresponds to the following two-qubit gate in the $ab =
\{00, 01, 10, 11 \}$ basis,
\begin{eqnarray}
U^{gate}_{k=3} = \left(\begin{array}{cccc} e^{i\alpha} & 0 & 0 & 0 \\
0 & e^{i\beta} & 0 & 0 \\
0 & 0 & e^{i(\alpha-\beta)} & 0 \\
0 & 0 & 0 & e^{-i\alpha}
\end{array}\right).\label{ukeq3}
\end{eqnarray}
As for $U_{k>3}^{gate}$, the $\beta$ dependence of $U_{k=3}^{gate}$
corresponds to a tensor product of single-qubit rotations. Gates of
this form with fixed $\alpha$ but different values of $\beta$ are
thus equivalent up to single-qubit rotations. If we use this
equivalence to set $\beta = \alpha$ we see that gates of the form
$U_{k=3}^{gate}$ are equivalent to the gates produced by our $F$
weave construction (\ref{fgate}), and so, in particular, when
$\alpha = \pi$ the resulting gate is equivalent to a controlled-NOT
gate.

\section{Conclusions}

To summarize, we have shown how to construct both single-qubit and
two-qubit gates for qubits encoded using nonabelian quasiparticles
described by $SU(2)_3$ CSW theory, or, equivalently, the $SO(3)_3$
theory (Fibonacci anyons).  Qubits are encoded into triplets of
quasiparticles and single-qubit gates are carried out by braiding
quasiparticles within qubits. Two classes of two-qubit gate
constructions were presented. In the first, a pair of
quasiparticles from one qubit is woven through those forming the
second qubit.  In the second, a single quasiparticle is woven
through three static quasiparticles (one from the same qubit as
the mobile quasiparticle, the other two from the second qubit). A
central theme in all of our two-qubit gate constructions is that
of breaking the problem of compiling braids for the six
quasiparticles used to encode two qubits into a series of braids
involving only three objects at a time.  While these constructions
do not in general produce the optimal braid of a given length
which approximates a desired two-qubit gate, we believe they do
lead to the most accurate (or at least among the most accurate)
two-qubit gates which can be obtained for a fixed amount of
classical computing power. Finally, we proved a theorem which
states that for the $SU(2)_k$ CSW theory, two-qubit gates
constructed by braiding only four quasiparticles (two from each
qubit) can only lead to leakage free entangling two-qubit gates
when $k=3$.

\acknowledgements

L.H.\ and N.E.B.\ acknowledge support from the US D.O.E.\ through
Grant No.\ DE-FG02-97ER45639, and G.Z. acknowledges support from
the NHMFL at Florida State University. L.H., N.E.B.\ and S.H.S.\
would like to thank the Kavli Institute for Theoretical Physics at
UCSB for its hospitality during which a major portion of this work
was carried out.


\end{document}